\documentclass{aa}  
\usepackage[table]{xcolor}
\usepackage{graphicx}
\usepackage{natbib,twoopt}
\bibpunct{(}{)}{;}{a}{}{,} 
\usepackage{txfonts}
\usepackage[colorlinks=true, linkcolor=black, citecolor=blue, urlcolor=blue, breaklinks=true]{hyperref}
\def\aap{A\&A}

\def\apjs{ApJS}
\def\apjl{ApJL}

\usepackage{amsmath,amssymb}
\usepackage{textcomp}
\usepackage{appendix}
\usepackage{xurl}
\usepackage{amsmath}
\usepackage{multirow}
\usepackage{booktabs}
\usepackage{rotating}
\usepackage{float}
\usepackage{placeins}
\usepackage{pdflscape}
\usepackage{stfloats}
\DeclareMathAlphabet{\pazocal}{OMS}{zplm}{m}{n}

\usepackage{siunitx}
\newcolumntype{T}[1]{S[table-format=#1]}
\newcolumntype{U}[1]{S[table-format=#1,
                    round-mode=places,
                    round-precision=2]}

\newcommand{\Mj}{M$_{\rm Jup}$}
\newcommand{\Av}{A$_\text{v}$}
\newcommand{\teff}{T\ensuremath{_{\text{eff}}}}

\newcommand{\Lagr}{\pazocal{L}}

\newcommand{\rep}[1]{\textcolor{orange}{(Repetition)}}

\begin{document}

\title{Disk fraction among free-floating planetary-mass objects in Upper Scorpius}

\author{T. Rodrigues\inst{1} \and
      H. Bouy\inst{1,2} \and 
      S. N. Raymond\inst{1} \and
      E. L. Mart\'in\inst{3} \and
      E. Bertin\inst{4} \and
      J. Olivares\inst{5} \and
      D. Barrado\inst{6} \and
      N. Huélamo\inst{6} \and
      M. Tamura\inst{7,8,9} \and
      N. Miret Roig\inst{10, 11} \and
      P. A. B. Galli\inst{12} \and
      J.-C. Cuillandre\inst{4}
      }

\institute{Laboratoire d'Astrophysique de Bordeaux, Univ. de Bordeaux, CNRS, B18N, Allée Geoffroy Saint-Hilaire, 33615 Pessac, France\\
          \email{tommy.rodrigues@u-bordeaux.fr, herve.bouy@u-bordeaux.fr}
         \and
         Institut universitaire de France (IUF), 1 rue Descartes, 75231 Paris CEDEX 05
         \and
         Instituto de Astrof{\'{\i}}sica de Canarias, Calle Vía Lactea s/n, E-38205 La Laguna, Tenerife, Spain
         \and
         Universit\'e Paris-Saclay, Universit\'e Paris Cit\'e, CEA, CNRS, 91191 AIM, Gif-sur-Yvette, France
         \and
         Departamento de Inteligencia Artificial, Universidad Nacional de Educación a Distancia (UNED), c/Juan del Rosal 16, E-28040, Madrid, Spain
         \and
         Centro de Astrobiología (CAB), CSIC-INTA, ESAC Campus, Camino bajo del Castillo s/n, Villanueva de la Ca\~nada, E-28692, Madrid, Spain
         \and
         Department of Astronomy, Graduate School of Science, The University of Tokyo, 7-3-1 Hongo, Bunkyo-ku, Tokyo, 113-0033, Japan
         \and
         Astrobiology Center, 2-21-1 Osawa, Mitaka, Tokyo, 181-8588, Japan
         \and
         National Astronomical Observatory of Japan, 2-21-1 Osawa, Mitaka, Tokyo, 181-8588, Japan
         \and
         Departamento de Física Quàntica i Astrofísica (FQA), Univ. de Barcelona (UB), Martí i Franquès, 1, 08028 Barcelona, Spain
         \and
         Institut de Ciències del Cosmos (ICCUB), Univ. de Barcelona (UB), Martí i Franquès, 1, 08028 Barcelona, Spain
         \and
         Instituto de Astronomia, Geofísica e Ciências Atmosféricas, Universidade de São Paulo, Rua do Matão, 1226, Cidade Universitária, 05508-090 São Paulo, SP, Brazil
          }

\date{Received July 7, 2025; accepted October 3, 2025}

\abstract
{Free-floating planetary-mass objects (FFPs) have been detected through direct imaging in several young, nearby star-forming regions. The properties of circumstellar disks around these objects may provide a valuable probe into their origin but are currently limited by the small sample sizes explored.}
{We aim to perform a statistical study of the occurrence of circumstellar disks down to the planetary-mass regime.}
{We performed a systematic survey of disks among the population identified in the 5–10\:Myr-old Upper Scorpius association, restricted to members outside the younger, embedded Ophiuchus region, and with estimated masses below 105\:\Mj. We took advantage of {\it unWISE} photometry to search for mid-infrared excesses in the {\it WISE} (W1--W2) color. We implemented a Bayesian outlier detection method, which models the photospheric sequence and computes excess probabilities for each object, enabling a statistically sound estimation of disk fractions.}
{We explored disk fractions across an unprecedentedly fine mass grid, reaching down to objects as low as $\sim$6\:\Mj\ assuming 5\:Myr or $\sim$8\:\Mj\ assuming 10\:Myr, thus extending the previous lower boundary of disk fraction studies. Depending on the age, our sample includes between 17 and 40 FFPs. We confirm that the disk fraction steadily rises with decreasing mass and exceeds 30\% near the substellar-to-planetary mass boundary at $\sim$13\:\Mj. We find hints of a possible flattening in this trend around 25\,--\,45\:\Mj, potentially signaling a transition in the dominant formation processes. This shift in trend should be considered with caution and needs to be confirmed with more sensitive observations. Our results are consistent with the gradual dispersal of disks over time, as disk fractions in Upper Scorpius appear systematically lower than those in younger regions.}
{}


\maketitle

\section{Introduction}

Free-floating planetary-mass objects (FFPs) roam the galaxy unbound to any star. Below the deuterium-fusion mass limit ($\lesssim$\,13\:\Mj), FFPs lack sufficient mass to sustain any nuclear fusion and thus continuously cool and fade over time \citep{Burrows2001, Chabrier2005, Spiegel2011}. Their very low luminosities and colors make them exceedingly difficult to detect and distinguish from background field or extra-galactic sources. Nonetheless, since their discovery, studies have found an increasing number of FFPs with masses down to just a few Jupiter masses in young, nearby star-forming regions, where they are still relatively warm and luminous at infrared (IR) wavelengths \citep{Tamura1998, Zapatero2000, Lucas&Roche2000, Barrado2001, PenaRamirez2012, Muzic2014, Muzic2015, Lodieu2018, Suarez2019, Luhman&Hapich2020, MiretRoig2022a, Langeveld2024, Martin2025, Luhman2024, DeFurio2025, Bouy2025}. Microlensing surveys have identified FFP candidates in the field with even lower masses, down to subterrestrial levels, as the method is sensitive to mass rather than luminosity \citep{Mroz2020, Sumi2023, Koshimoto2023, Jung2024}. However, it is important to note that their nature may not be genuinely free-floating, as microlensing surveys cannot distinguish them from wide-separation companions. The James Webb Space Telescope (JWST) has detected new populations of FFPs in nearby star-forming regions down to 3--4\:\Mj \citep{Luhman2024, Langeveld2024, DeFurio2025}. In NGC\:2024, \citet{DeFurio2025} report a hint of a turnover in the stellar initial mass function (IMF) below 12\:\Mj, with an apparent cutoff near 3\:\Mj despite deeper sensitivity, even though the candidates still lack spectroscopic confirmation. \citet{Langeveld2024} reported a similar cutoff around 5\:\Mj\ in NGC\:1333, with spectral types confirmed for the candidates. If confirmed, such a cutoff would align with theoretical predictions for the opacity limit to fragmentation \citep{Whitworth2024, Bate2012}, suggesting that there may be a universal threshold for the isolated formation of planetary-mass objects.

Free-floating planetary-mass objects may have formed either as planets originally bound to stars or through a mechanism similar to that of stars \citep{MiretRoig2023}. Similar to more massive brown dwarfs (BDs), four main scenarios have been proposed to explain FFP formation: (1) formation within a star's protoplanetary disk either by core accretion \citep{Pollack1996} or gravitational fragmentation \citep{Boss1998}, followed by ejection from the parent system by planet-planet scattering \citep{Chatterjee2008, Boley2012, Juric&Tremaine2008, Veras&Raymond2012, Daffern-Powell2022}, close stellar encounters \citep{Parker&Quanz2012, Yu2024, Wang2015, Wang2024}, or ejection from circumbinary systems \citep{Chen2024, Coleman2024}; (2) scaled-down star formation via direct core-collapse and turbulent fragmentation of small molecular clouds \citep{Bate2002, Padoan&Nordlund2004, Hennebelle&Chabrier2008}; (3) as ejected stellar embryos from dense stellar nurseries \citep{Reipurth&Clarke2001}; and (4) by photo-erosion of pre-stellar cores due to nearby OB stars, although this should produce only a few objects \citep{Whitworth&Zinnecker2004}. 
The observed excess of FFPs compared to core-collapse model predictions suggests that several of these different pathways likely contribute to FFP formation \citep{Bouy2009, PenaRamirez2012, Muzic2019, MiretRoig2022a, Langeveld2024}. Further studies are required to quantify the role of each in determining the final FFPs population and their dependence on the environment.

Soon after their discovery, some FFPs were found to host accretion disks \citep{Testi2002, Barrado2002, Jayawardhana2006, Luhman2005b, Luhman2005a}, similar to young stars and BDs. While disks are likely to be found at early times in all the formation scenarios mentioned above, their survival depends on subsequent dynamics; for example, the ejection of BDs or planets can truncate disks and sometimes even strip them entirely \citep{Umbreit2011, Testi2016, Rabago&Steffen2019}. Measuring the occurrence rate of disks around FFPs and understanding their physical properties are therefore essential for improving our understanding of the diversity and evolution of BDs and FFPs as well as their respective planet and/or moon-forming environments.
Recent studies have begun to probe disk fraction within the planetary-mass regime in a few nearby star-forming regions. In NGC\:1333, \citet{Scholz2023} reported a disk fraction of $42^{+18}_{-16}\%$ among 12 spectroscopically confirmed objects with spectral types M9--L3, corresponding roughly to 6--12\:\Mj. Similarly, \citet{Seo&Scholz2025} found a disk fraction of $46^{+13}_{-12}\%$ for M9--L1 (9--13\:\Mj) objects in IC\:348. Other surveys have also reported significant disk fractions extending into the planetary-mass regime in Taurus, NGC\:1333, IC\:348, and Upper Sco \citep{Luhman2016, Esplin&Luhman2019, Luhman&Esplin2020, Luhman2025}, although with coarser mass bins. Unfortunately, constraints on disk fraction among FFPs are currently limited by the large uncertainties associated with small sample sizes \citep[typically a dozen at most, ][]{Scholz2023, Seo&Scholz2025}, as well as issues of incompleteness and contamination \citep{Allers2006, Zapatero2007, PenaRamirez2012, Esplin&Luhman2019, Scholz2023}.

In this paper, we present a comprehensive study of the disk fraction around FFPs and BDs within the young \citep[5\,--\,10\:Myr,][]{David2019, Pecaut2012, Pecaut&Mamajek2016, MiretRoig2022b, Ratzenbock2023}, nearby \citep[120\,--\,145\:pc,][]{deBruijne1997, Fang2017} Upper Scorpius (USC) OB association. The recently identified large sample of coeval FFPs (70\,--\,170 candidates, depending on the assumed age) down to $\sim$4\:\Mj\ \citep{MiretRoig2022a}, with low contamination levels \citep[$\leq$6\%, ][]{Bouy2022}, provides an excellent basis for a statistical study of disk occurrence among FFPs. The region has been observed extensively by {\it Wide-field Infrared Survey Explorer} \citep[{\it WISE};][]{Wright2010}, providing deep mid-IR data ideally suited to identify the presence of hot circumstellar material. 

The paper is structured as follows. In Sect.~\ref{sect:sample} we describe the working sample in more detail; in Sect.~\ref{sect:method} we describe our method for identifying IR excesses that indicate the presence of a disk; in Sect.~\ref{sect:result}, we present the computed disk fraction estimates and, in Sect.~\ref{sect:discussion}, we discuss their implications for the origin of FFPs and compare our findings with the existing literature; finally, we summarize our conclusions in Sect.~\ref{sect:conclusion}.

\section{Sample}
\label{sect:sample}

The large population reported in \cite{MiretRoig2022a} includes 3\,455 high-probability candidate members of the USC and Ophiuchus (Oph) star-forming regions, among which between 70 and 170 are identified as FFPs, depending on the age assumed, down to masses as low as $\sim$4\:\Mj. This dataset therefore constitutes an excellent laboratory to investigate the disk population of FFPs. To assess the reliability of the membership, \cite{Bouy2022} examined independent spectroscopic youth indicators for a subset of 17 FFP candidates and found a low contamination rate ($\leq6\%$) among USC members. In the present work, we focused on mid-IR excess detection (Section~\ref{sect:method}) for the subsample with estimated masses below 105\:\Mj. This upper mass limit was chosen to include the star-brown dwarf and brown dwarf-planetary mass objects transitions, enabling both meaningful comparisons with existing literature and the inclusion of enough objects to provide statistically significant results. 

\subsection{Age and mass}
\label{sect:age}

The area covered by \cite{MiretRoig2022a} in the Scorpius star-forming complex includes various subgroups with diverse ages and evolutionary stages, from the very young Oph clouds \citep[1\,--\,3\:Myr,][]{Greene&Meyer1995, MiretRoig2022b} to the older populations in the region known as USC. Several studies have estimated the age of USC between 5 and 10\:Myr \citep{David2019, Pecaut2012, Pecaut&Mamajek2016, MiretRoig2022b, Ratzenbock2023}. This age spread is largely due to overlapping populations within the region, as demonstrated by 3D spatial distributions and kinematics analyses \citep{Kerr2021, MiretRoig2022b, Ratzenbock2023}. However, disentangling these subgroups and assigning corresponding ages to individual objects remains challenging, especially for the BD and FFP members lacking parallax and radial velocity measurements. In the following, we carry out the analysis and present the results obtained for both 5 and 10\:Myr, using the masses estimated by \cite{MiretRoig2022b} for these two ages. Masses were inferred simultaneously with extinctions using the {\tt Sakam}\footnote{\url{https://github.com/olivares-j/Sakam}} Bayesian framework \cite{Olivares2019}, by comparing the absolute magnitudes with the BHAC15 \citep{Baraffe2015} evolutionary models. These mass estimates assume distances inferred from Gaia data release 2 (DR2) \citep{GaiaDR2} parallaxes using {\it Kalkayotl}\footnote{\url{https://github.com/olivares-j/kalkayotl}} \citep{Kalkayotl} with a Gaussian prior centered at $145\pm45$\:pc. For sources without parallaxes, distances were sampled from the inferred cluster distribution \citep{MiretRoig2022b}.

\subsection{Area selection}
\label{sect:area}

\begin{figure}[h]
\centering
\includegraphics[width=\columnwidth]{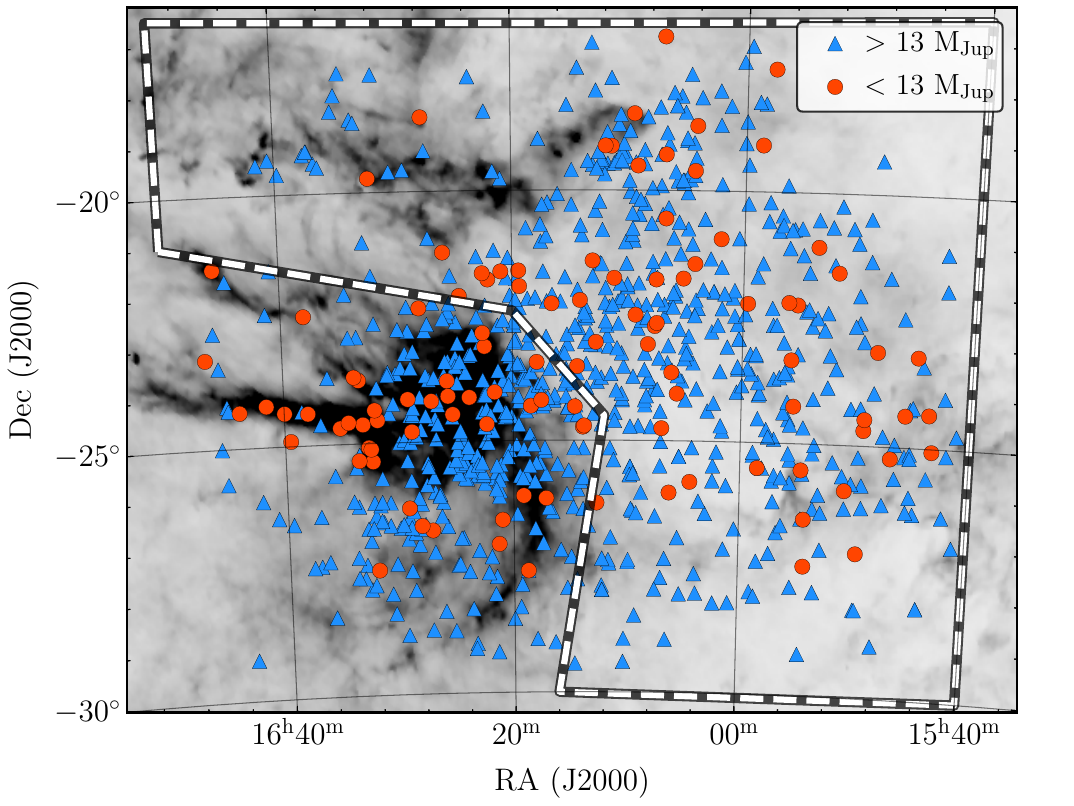}
\caption{Spatial distribution of USC and Oph members from \cite{MiretRoig2022a}. Objects with estimated masses below 13\:\Mj\ and between 13\,--\,105\:\Mj\ (for an assumed age of 5\:Myr) are represented by red circles and blue triangles, respectively. The dashed white line indicates the extinction boundary used in this study, as defined by the polygon in Sect.~\ref{sect:area}. Background image: Planck 857\:GHz \citep{Planck}.}
\label{spatial_map}
\end{figure}

The completeness and misclassification rates of the FFP sample in regions such as Oph, the youngest and most embedded region in Scorpius, are expected to worsen given the large extinction levels and the degeneracy between the effective temperature (\teff) and extinction \citep{Bailer-Jones2011, MiretRoig2022a}. In the present study, we therefore chose to exclude the Oph region and restrict our analysis to areas of low to moderate extinction (\Av\,$\lesssim$\,1\,mag). Boundaries were heuristically defined using the Planck 857 GHz dust emission map \citep{Planck}, a reliable proxy for line-of-sight extinction, and illustrated in Fig.~\ref{spatial_map}.
The selected region encompasses $\sim$155 square degrees in USC and is outlined as a polygon with the following vertices, expressed in J2000 equatorial coordinates (RA, Dec) in degrees: (252.5, -21), (252.5, -16.5), (235, -16.5), (235, -30), (244, -30), (243, -24.5), and (245, -22.4). This area contains 528 objects, including between 221 and 378 BDs, and between 32 and 61 FFPs, depending on whether an age of 10 or 5\:Myr is assumed.

\subsection{Working sample}
\label{sect:final_sample}

In parallel, we excluded 25 sources from the analysis for various reasons. 
As Gaia data release 3 (DR3) \citep{GaiaDR3} parallaxes are now available for these objects, Ten, initially selected as candidate members from the Dynamical Analysis of Nearby ClustErs (DANCe) sample without parallax constraints \citep{MiretRoig2022b}, were later found to lie at distances exceeding 300\:pc. This is incompatible with the USC association’s location at approximately 145\:pc \citep{Fang2017}, indicating that these are likely background contaminants. They are therefore excluded from our final sample.
Among these ten sources, between zero (at 10\:Myr) and nine (at 5\:Myr) were previously classified as BDs assuming a USC distance.
In 15 additional cases, the absence of photometric data in one or both of the {\it WISE} W1 and W2 mid-IR bands prevents us from assessing the color used to identify excesses, as explained in Sect.~\ref{sect:method}. These 15 objects include in particular one FFP and between 11 and 14 BDs.
After discarding these sources, the final working sample consists of 503 members (given in Table~\ref{tab:members_sample}), representing 95\% of the original population below 105\:\Mj\ in USC. Depending on the assumed age, it includes between 31 and 60 objects below 13\:\Mj, and between 210 and 355 objects in the 13--75\:\Mj\ range.

\section{Circumstellar disk identification}
\label{sect:method}

Circumstellar disks around young stellar objects are typically identified by their excess emission at IR wavelengths, produced when dust in the disk absorbs and reemits stellar radiation at mid- and far-IR wavelengths. A number of methods have been used to identify IR excesses, often using IR colors and IR spectral slopes ($\alpha$), or by making comparison with photospheric models \citep[see for example][ and references therein]{Allen2007, Lada2006, Teixeira2020, Luhman2008, Koenig&Leisawitz2014, Ribas2015, Esplin2018, Luhman&Mamajek2012}.

\subsection{Infrared photometry}
\label{sect:photometry}

We took advantage of the sensitivity and homogeneous coverage of {\it WISE} \citep{Wright2010} to look for excess between 3.4\:$\mu$m (W1 band) and 4.6\:$\mu$m (W2 band). The (W1--W2) color has been indeed successfully used in previous studies to detect IR excess among stars, BDs, and FFPs \citep[e.g.,][]{Dawson2013, Riaz2012, Luhman2023b, Luhman2025}. Because these wavelengths probe the hot inner regions of the disk, excesses in these bands are indicative of full, optically thick (primordial) disks (class II). {\it WISE} also measured the luminosity at two longer wavelengths—the W3 (12\:$\mu$m) and W4 (22\:$\mu$m) bands—which are also valuable for detecting disk excess emission, especially when the disks are more evolved. However, these two bands are significantly shallower and do not reach the planetary-mass regime. For this reason, we restricted our analysis to the W1 and W2 bands.

\begin{figure}[h]
\centering
\includegraphics[width=\columnwidth]{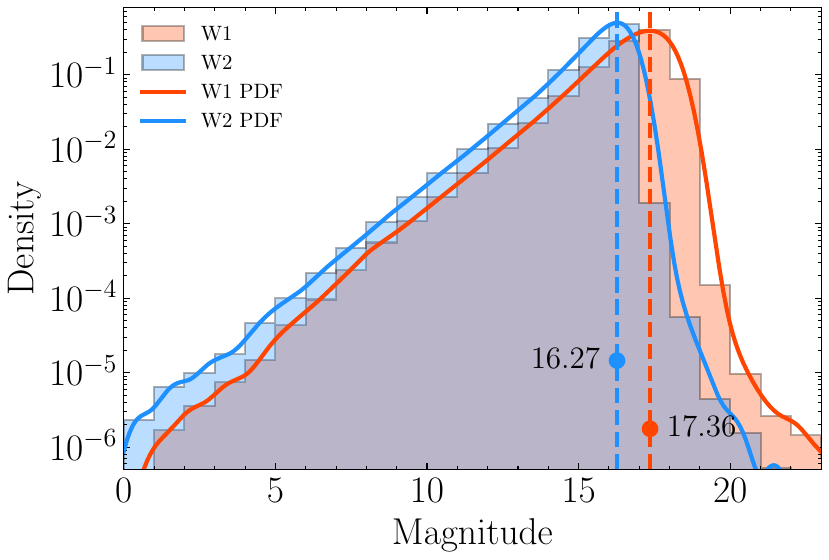}
\caption{Source density as a function of magnitude in the {\it unWISE}/W1 and {\it unWISE}/W2 bands. Solid lines represent the kernel density estimates of the histograms, while vertical dashed lines mark the turning points used to define the completeness limits of the {\it unWISE} data.}
\label{fig:histo_completeness}
\end{figure}

We cross-matched the sample (described in Sect.~\ref{sect:sample}) with the latest {\it unWISE} catalog \citep{unWISE, Meisner2021}, which is built from the co-addition of all publicly available W1 and W2 {\it WISE} images and provides homogeneous coverage across the USC region. The photometry is shown in Table~\ref{tab:members_sample}. Uncertainties on the (W1--W2) color were derived through standard error propagation as $\sigma_{\mathrm{W1-W2}} = \sqrt{\sigma_{\mathrm{W1}}^2 + \sigma_{\mathrm{W2}}^2}$, where $\sigma_{\mathrm{W1}}$ and $\sigma_{\mathrm{W2}}$ are the photometric uncertainties in W1 and W2, respectively.

The completeness limits of the {\it unWISE} catalog in the W1 and W2 bands were estimated by examining the magnitude distribution of sources located in low-extinction regions of the field of view, within the polygon defined in Sect.~\ref{sect:area}. The peak of this distribution served as an estimate of the completeness limit, identified as W1$_{\rm comp}\approx$17.36\,mag and W2$_{\rm comp}\approx$16.27\,mag (see Fig.~\ref{fig:histo_completeness}).
When translated into mass limits using the evolutionary models of \citet{Baraffe2015}, and assuming a median distance of 145\:pc \citep{deBruijne1997, Fang2017}, this corresponds to $\leq$4\:\Mj\ at 5\:Myr and $\sim$6\:\Mj\ at 10\:Myr for W1$_{\rm comp}$, and $\sim$6\:\Mj\ at 5\:Myr and $\sim$8\:\Mj\ at 10\:Myr for W2$_{\rm comp}$.
These mass limits are applicable to sources without IR excess and help define the minimum mass down to which diskless members can be reliably detected and included in our analysis. Since both W1 and W2 detections are required to compute the color used for identifying excesses, we adopted the W2 completeness limit—which corresponds to the shallower band—as the lower mass boundary for computing the disk fractions. This ensures that the disk fractions are not biased in favor of disk-bearing sources due to incompleteness among diskless objects.

To minimize false excess detections, we performed a careful visual inspection of the {\it unWISE} images for all objects in our sample. This was supplemented by higher-resolution ground-based optical and near-IR imaging from the \citet{MiretRoig2022a} survey to identify potentially unreliable photometric measurements. The latter seeing-limited ground-based images have a typical resolution of $\sim$1\arcsec. We adopted a conservative approach, flagging any sources that showed evidence of blending or binarity in either the optical or IR images. This strategy prioritized reliability over completeness, as unresolved nearby sources can artificially inflate mid-IR flux and mimic disk-related excess. In particular, any object with a nearby source in the optical or near-IR images falling within the {\it unWISE} point spread function (PSF) was flagged as suspicious. Since the occurrence of blending is expected to be random, this filtering process should not introduce any systematic bias in our analysis of disk fractions. A total of 94 out of 503 sources (18\% of the sample) were flagged as blended, including between 14 and 20 objects below 13\:\Mj, depending on the assumed age. In the following, we present and discuss our results based on the filtered sample of 409 unflagged sources, which contains between 17 and 40 objects below 13\:\Mj. Being 140--330\% larger—depending on the assumed age—than the largest samples used to derive disk fractions among FFPs to date, as reported by \citet{Scholz2023} and \citet{Seo&Scholz2025}, this sample represents a significant improvement over previous efforts.

\subsection{Method to detect infrared color excesses}
\label{sect:MCMC}

\begin{figure*}[h]
\centering
\includegraphics[width=\textwidth]{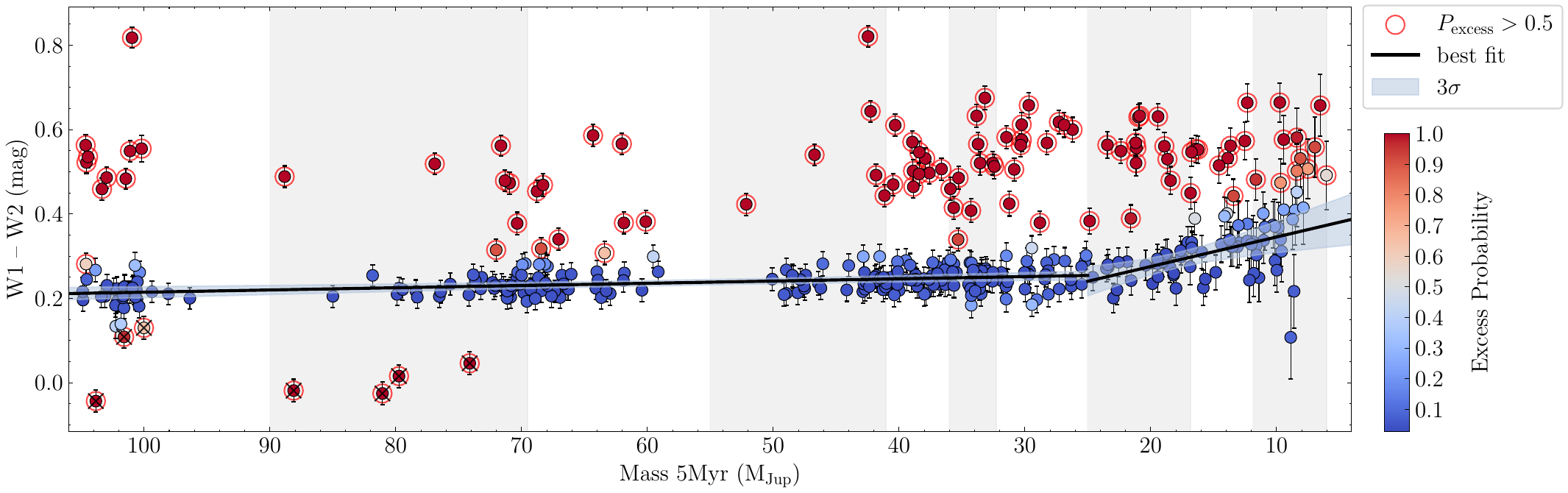}
\includegraphics[width=\textwidth]{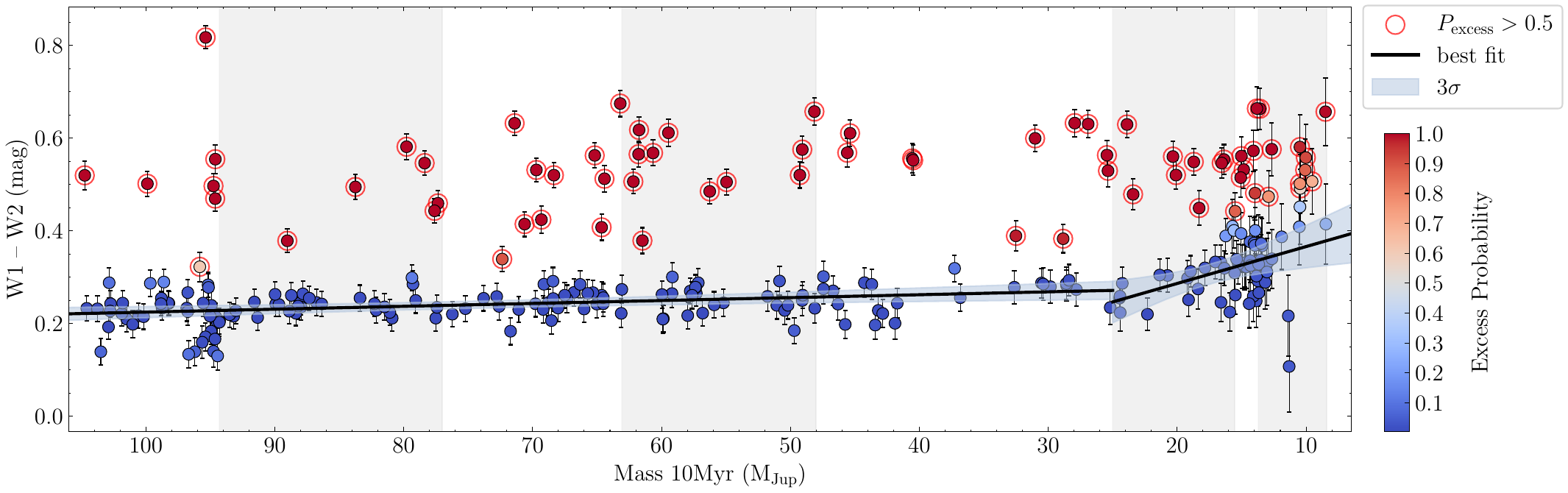}
\caption{W1--W2 IR color as a function of mass for USC members, assuming ages of 5 ({\it Top}) and 10\:Myr ({\it Bottom}). The solid black line represents the best-fit relation, while the blue-shaded region corresponds to the 3$\sigma$ confidence interval of the fit parameters (i.e., slope and intercept). The color scale represents the inferred excess probability, as indicated by the color bar. Red circles highlight sources with an excess probability greater than 0.5. Objects identified as outliers ($P_{\mathrm{excess}}>0.5$) below the fitted photospheric sequence are marked with black crosses, as they are excluded from the subsequent analysis. Background shading is used to delineate mass bins used in Figures~\ref{fig:violin_fractions_woWbl} and \ref{fig:fractions_lit}.}
\label{fig:MCMC_plot_woWbl}
\end{figure*}

\begin{figure*}[h]
\centering
\includegraphics[width=\textwidth]{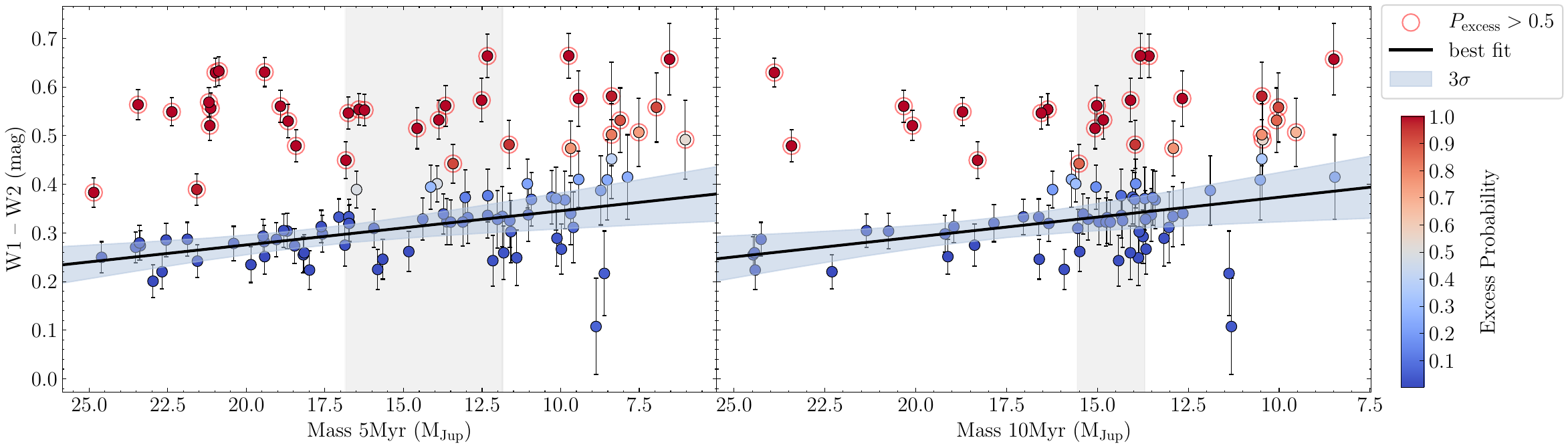}
\caption{W1--W2 IR color as a function of mass for USC members in the 0--25\:\Mj\ mass range, assuming ages of 5 and 10\:Myr respectively. This zoomed-in view focuses on the low-mass regime to better visualize the linear fits and the associated excess probabilities in this domain, shown in Fig.~\ref{fig:MCMC_plot_woWbl} and discussed in Sect.~\ref{sect:MCMC}. See the caption in Fig.~\ref{fig:MCMC_plot_woWbl} for a description.}
\label{fig:MCMC_plot_lows_woWbl}
\end{figure*}

A proven method to detect disk-related IR excesses consists in analyzing observed IR colors in relation to the spectral type \citep{Luhman&Esplin2020, Esplin&Luhman2019}. In this context, objects with circumstellar disks appear redder than their diskless counterparts due to the excess IR emission originating from the disk. Notably, because colors are relative measurements, they are largely unaffected by extinction and distance. This makes them a robust tracer of IR excess when compared across objects with similar intrinsic properties (e.g., spectral type, luminosity, or effective temperature).
Unfortunately, most sources in the sample lack spectroscopic observations and do not have measured spectral types. To address this limitation, we used the mass inferred by \cite{MiretRoig2022a} (see Sect.~\ref{sect:age}), which acts as a proxy for intrinsic properties, instead of spectral types.

Figure~\ref{fig:MCMC_plot_woWbl} presents the (W1--W2) color as a function of estimated mass at ages of 5 and 10\:Myr. Sources without IR excess align along a well-defined sequence representing photospheric emission, which remains particularly distinct down to $\sim$25\:\Mj. As noted by \citet{Luhman&Mamajek2012}, \citet{Luhman&Esplin2020}, and \citet{Esplin&Luhman2019}, the slope of this sequence appears to change below $\sim$25\:\Mj, indicating that a different photospheric color-mass relation may be required in this regime. A handful of obvious outliers located below the fitted photospheric sequence (see upper panel of Fig.~\ref{fig:MCMC_plot_woWbl}) likely corresponds to more massive objects whose masses were underestimated when using the 5\:Myr isochrone \citep{MiretRoig2022b}. They were excluded from the subsequent analysis.

\renewcommand{\arraystretch}{1.4}
\begin{table*}
\centering
\caption{Posterior estimates of the key regression and outlier model parameters from our Bayesian inference.}
\label{tab:MCMC_results_woWbl}
\begin{tabular}{lcccc}
\toprule
& \multicolumn{2}{c}{\textbf{Mass 5 Myr}} & \multicolumn{2}{c}{\textbf{Mass 10 Myr}} \\
\cmidrule(lr){2-3} \cmidrule(lr){4-5}
                           Parameters & \hspace{2mm}$[25$--$105]$\:\Mj\hspace{3mm} & \hspace{3mm}$[0$--$25]$\:\Mj \hspace{3mm} 
                            & \hspace{3mm}$[25$--$105]$\:\Mj \hspace{3mm} & \hspace{3mm}$[0$--$25]$\:\Mj \vspace{1mm} \\ 
\midrule
$\text{intercept}$ & $0.267 _{-0.011}^{+0.009}$    & $0.415_{-0.050}^{+0.045}$     & $0.287_{-0.018}^{+0.017}$     & $0.446_{-0.06}^{+0.061}$      \\
$\text{slope}$     & $-0.0005_{-0.0002}^{-0.0002}$ & $-0.0070_{-0.0025}^{+0.0027}$ & $-0.0006_{-0.0002}^{+0.0002}$ & $-0.0080_{-0.0035}^{+0.0032}$ \\
\addlinespace
$P_b$              & $0.290_{-0.055}^{+0.057}$     & $0.363_{-0.101}^{+0.103}$     & $0.252_{-0.063}^{+0.069}$     & $0.329_{-0.117}^{+0.125}$     \\
$Y_b$              & $0.430_{-0.042}^{+0.041}$     & $0.526_{-0.035}^{+0.034}$     & $0.514_{-0.045}^{+0.041}$     & $0.531_{-0.044}^{+0.039}$     \\
$\sigma_b$         & $0.179_{-0.025}^{+0.028}$     & $0.073_{-0.023}^{+0.036}$     & $0.106_{-0.028}^{+0.047}$     & $0.056_{-0.027}^{+0.055}$     \\
\bottomrule
\end{tabular}
\tablefoot{
We separately show results for the high-mass ($25-105$\:\Mj) and low-mass ($0-25$\:\Mj) regimes and for the analysis with masses assuming ages of 5 and 10\:Myr. The mass division between the low- and high-mass regimes is set at $M = 25$\:\Mj, following the observed break in the sequence (see Sect.~\ref{sect:MCMC}). Reported values correspond to the posterior median, with uncertainties given as 3--97\% HDIs in subscript and superscript.
$P_b$ is the global outlier fraction in the population. $Y_b$ and $\sigma_b$ are the center and scale of the outlier (excess) distribution in W1--W2 color space. The slope and intercept correspond to the best-fit linear regression for the photospheric sequence. Posteriors are well-converged with $\hat{R} \simeq 1.0$ for all parameters.}
\end{table*}

In this study, we adopted a statistical framework to estimate the probability that an object with a given mass exhibits an excess in (W1--W2) indicative of circumstellar disk emission. We fit a line to the data presented in Fig.~\ref{fig:MCMC_plot_woWbl} using the Bayesian outlier detection method described by \citet{Hogg2010}, employing Markov chain Monte Carlo (MCMC) techniques to identify outliers--defined here as sources with excess (W1--W2) color due to the presence of a disk. A different line was independently fit for the mass domain M$\ge$25\:\Mj\ and the mass domain M$<$25\:\Mj. In the model, each point $i$ had an associated latent binary variable $q_i$, sampled at each iteration: $q_i=1$ for inliers and $q_i=0$ for outliers, indicating whether the point is consistent with the fitted photospheric sequence. These were modeled as drawn from a Bernoulli distribution with parameter $1 - P_b$, where $P_b$, the global outlier fraction inferred by the model, represents the probability that any given data point is an outlier. At each iteration, only the inliers contribute to the likelihood of the regression fit. By contrast, the outliers are rejected and instead modeled by a broader distribution centered on $Y_b$, with a spread $\sigma_b$ (both in W1--W2 color), and governed by $P_b$. We adopted uniform priors for these parameters to allow flexibility when modeling the outlier population without imposing strong constraints. The MCMC sampling was performed over two chains of 5\,000 samples each (a total of 10\,000 samples), across which the model computed the total log-likelihood by summing the contribution from the inlier and outlier likelihoods for each point, simply expressed as

\begin{equation}
\hspace{0.08\textwidth}
\begin{aligned}
\log \Lagr & = \sum_{i}  \left[q_i \log p_{\text{in},i} + (1 - q_i) \log p_{\text{out},i} \right]\\
& = \sum_{i} \left[\log \Lagr_{\text{in},i} + \log \Lagr_{\text{out},i} \right]
\label{eq:outlier_likelihood},
\end{aligned}
\end{equation}

\noindent where $L_{\text{in},i}$ and $L_{\text{out},i}$ represent the likelihood of point $i$ being an inlier or outlier, and $p_{\text{in},i}$ and $p_{\text{out},i}$ denote the generative models for inliers and outliers, respectively. This sampling yields posterior probability distributions for all model parameters, which are summarized in Table~\ref{tab:MCMC_results_woWbl}. The median and the 3--97\% highest density credible intervals (HDI) for the posterior distributions of the slopes and intercepts of the 5 and 10\:Myr fits are listed for both the low- and high-mass regimes, along with the parameters of the outlier models ($Y_b$, $\sigma_b$, and $P_b$).
As this table indicates, the inferred fit parameters for the two ages are well consistent within the estimated uncertainties across both mass regimes. A similar agreement is observed for the parameters of the outlier models.
The resulting distinct linear fits for the mass domain M$\ge$25\:\Mj\ and the mass domain M$<$25\:\Mj\ are in good agreement and intersect at 25\:\Mj\ with only minor offsets of 0.012 and 0.031\:mag for 5\:Myr and 10\:Myr, respectively, which are included to a large extent within the scatter of the fit (see Fig.\ref{fig:MCMC_plot_woWbl}).
A zoom-in on the low-mass domain (M$\leq$25\:\Mj) is shown in Fig.~\ref{fig:MCMC_plot_lows_woWbl}. 
We carefully verified the convergence of the MCMC chains by inspecting the trace plots for all parameters, all of which exhibit good stationarity. In addition, the potential scale reduction factor $\hat{R}$ \citep{Gelman&Rubin1992}, widely used as a convergence diagnostic, is $\hat{R}\simeq1$ for all parameters, confirming that all parameters are well-converged.

Compared to previous binary classification approaches based solely on color thresholds, this method provides a probabilistic excess assessment for each source, allowing for more robust estimates of disk fractions and their associated uncertainties.
In this study, we therefore assumed that the outlier probability provides a statistically robust quantitative measure of the likelihood that a given object exhibits IR excess, i.e., it lies significantly above the fitted photospheric sequence. For a given point $i$, this value was computed from the posterior distribution of $q_i$ via the quantity,

\begin{equation}
\begin{aligned}
P_{\mathrm{excess}, i} \equiv P_{\mathrm{outlier}, i} & = P\left(q_i=0\right) = 1-P\left(q_i=1\right) \\
& = 1-\langle q_i^{(k)}\rangle_k = \frac{1}{N_k} \sum_{k}^{N_k} \left(1 - q_i^{(k)}\right)
\label{eq:P_exc},
\end{aligned}
\end{equation}

\noindent where the brackets denote the mean value over all MCMC samples $k$ --which reflects the fraction of iterations in which the object is classified as exhibiting an excess-- $N_k$ the number of iterations, and $q_i^{(k)}$ the value of the latent variable $q_i$ at iteration $k$. A source is thus considered to host a disk whenever its (W1--W2) color both lies above the fitted photospheric sequence and exhibits sufficiently high excess probability, typically greater than 0.5. Table~\ref{tab:members_sample} presents the excess probabilities inferred for each object in our sample, based on both the 5 and 10\:Myr analyses.
The aforementioned outliers located below the photospheric sequence in Fig.~\ref{fig:MCMC_plot_woWbl} are marked with negative excess probability in Table~\ref{tab:members_sample} and are excluded from the disk fraction calculations described in Sect.~\ref{sect:result}.
Our implementation was carried out in Python, utilizing the {\tt astroML} \citep{astroML, astroMLText} and {\tt PyMC} \citep{pymc} libraries.

\subsection{Validation of the method}
\label{sect:validation}

To assess the reliability of our method, we compared our results with those from the study by \citet{Luhman&Esplin2020} in the USC region. Their work, based on the spectral type–color diagram method described in \citet{Luhman&Mamajek2012}, provides a useful comparison sample with spectroscopically determined spectral types. We cross-matched the two samples to identify common members and specifically compare their W2 excess detections with those identified in our analysis -- defined as sources with an excess probability greater than 0.5, indicating statistically significant excess.

\begin{figure}[h]
    \centering
    \begin{minipage}{0.49\linewidth}
        \centering
        \includegraphics[width=\linewidth]{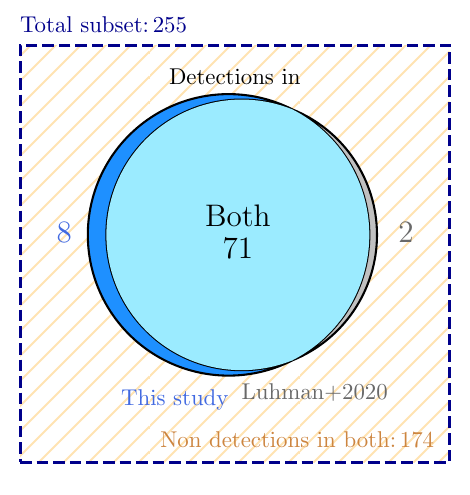}
    \end{minipage}
    \hfill
    \begin{minipage}{0.49\linewidth}
        \centering
        \includegraphics[width=\linewidth]{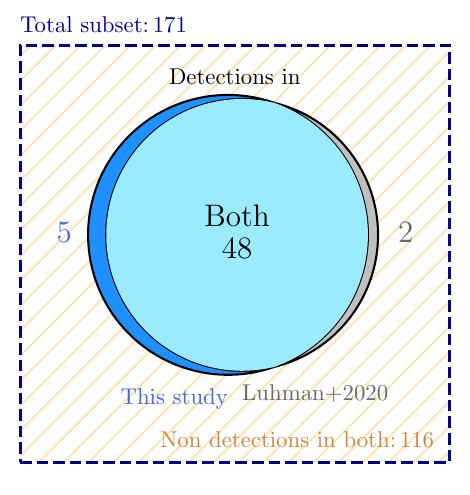}
    \end{minipage}
    \caption{Comparison of W2 excess detections between this study and \cite{Luhman&Esplin2020} using Venn diagrams, for assumed ages of 5\:Myr ({\it Left}) and 10\:Myr ({\it Right}). In each case, the central intersection shows the number of sources identified as exhibiting excess in both studies, while the left and right circle arcs show those detected only in one study but not the other. The orange-hatched region represents sources without excess detection in either work.}
    \label{fig:venn_w}
\end{figure}

Figure~\ref{fig:venn_w} presents a Venn diagram illustrating the overlap and the differences between the two sets of the W2 excess detections. The comparison is based on a common subset of 255 and 171 USC members with masses $\le$105\:\Mj, assuming ages of 5 and 10\:Myr, respectively, and located within our defined spatial boundaries (see Sect.~\ref{sect:area}). Extended versions of these diagrams--including sources flagged as potentially blended in the {\it unWISE} images--are provided in Fig.~\ref{fig:venn} in the Appendix, increasing the common sample to 300 and 207 members at 5 and 10\:Myr, respectively.

The discrepancies represent approximately 4\% of the total shared subset at both 5 and 10\:Myr, indicating strong overall agreement between the two methods.
Notably, we find that the agreement remains stable across a wide range of excess probability thresholds. The fraction of matching detections remains nearly unchanged up to a threshold of 0.95 and is maximized near 0.8, indicating insensitivity to variations in the excess probability threshold.

While \citet{Luhman&Esplin2020} based their analysis on the $K_s$--W2 color with W2 photometry from the {\it AllWISE} source catalog \citep{Cutri2013}, our study relies on the W1--W2 color, utilizing the significantly deeper and more recent {\it unWISE} catalog \citep{unWISE}. As a result, some of the observed discrepancies are attributed to differences in both the adopted photometry and the choice of reference band ($K_s$ vs. W1), particularly at fainter magnitudes. These variations reflect inherent observational limitations rather than fundamental issues with either method.

Methodological differences may nevertheless contribute to these small differences. In particular, our approach estimates empirical photospheric colors as a function of mass, whereas \citet{Luhman&Esplin2020} rely on spectral types. This distinction introduces additional sources of uncertainty. Nonetheless, the excellent level of agreement observed validates the methodology. 

\section{Disk fractions estimates}
\label{sect:result}

We derived the disk frequency—defined as the percentage of sources exhibiting IR excess relative to the total number of sources—as a function of the central object’s mass. The main analysis presented in this section is based on the sample excluding flagged sources in {\it unWISE} images (Sect.~\ref{sect:photometry}), while an equivalent analysis including them is provided in Appendix~\ref{app:analysis_including_flagged}. The linear regression provides an estimate of $P_b$, the global probability of exhibiting an excess, representing the overall fraction of sources with excesses across the dataset. However, because $P_b$ is a global parameter, it does not capture variations in excess frequency across different mass bins. To address this, we utilized the latent variable $q_i$ and the quantity $1-\langle q_i\rangle$, which represents the probability that each individual source exhibits an excess, thereby allowing us to estimate the excess (disk) fraction as a function of mass. For a given mass bin, the excess fraction $f^{(k)}$ (i.e., the fraction of $q_i=0$) at each sampling iteration $k$ was computed as

\begin{equation} 
\hspace{0.07\textwidth}
f^{(k)}_{\rm excess, bin} = \langle1-q_i^{(k)}\rangle_i = \frac{1}{N_i} \sum_{i \in bin}^{N_i} \left(1 - q_i^{(k)}\right)
\label{eq:frac_out}, 
\end{equation}

\noindent where $N_i$ denotes the number of objects in the bin and $q_i^{(k)}$ is the value of the latent variable $q_i$ for object $i$ at iteration $k$.
To estimate the disk fraction values in each mass bin, we used the median of the $\langle1-q_i^{(k)}\rangle_i$ posterior distribution, computed over all MCMC iterations $k$, as the central value. This approach naturally incorporates uncertainties from both the model parameters (the line representing the photospheric emission) and the individual classifications. These distributions are presented in Fig.~\ref{fig:violin_fractions_woWbl} and the corresponding derived fractions are listed in Appendix Table~\ref{tab:fractions}. We also deliberately kept the disk fractions computed on either side of the 25\:\Mj\ threshold -- used to separate the two distinct linear fits -- separate, to ensure consistency with the fit parameter dispersion.

To ensure statistically meaningful results, the disk fractions were computed within mass bins containing a minimum of 40 objects, with a target of 50 objects per bin whenever possible. This choice provides a robust sampling of the mass function, balancing statistical significance with sufficient mass resolution—particularly in the low-mass regime.
At higher masses, larger bin sizes were occasionally adopted to maintain a uniform distribution of sources across mass intervals, as illustrated by the point density overlay in Fig.~\ref{fig:violin_fractions_woWbl}.

We derived the associated uncertainties from the 68\% and 95\% confidence intervals (i.e., 1$\sigma$ and 2$\sigma$), calculated from the quantiles of the $\langle1-q_i\rangle_i$ distribution.
This method provides a statistically robust error estimate compared to the standard binomial approximations often adopted in disk fraction studies \citep[e.g.,][]{Scholz2023, Luhman&Esplin2020, Damian2023, Seo&Scholz2025}, is expected to capture most error sources, and robustly quantify confidence levels for the disk fractions.
In certain mass bins, the derived disk fraction exhibits minimal variation across iterations. This behavior arises when the majority of excess-bearing sources are clearly separated from the photospheric sequence, resulting in a low likelihood of misclassification as excesses. When combined with a well-defined photospheric locus, this yields highly stable $q_i$ values across iterations and thus a narrow distribution of $\langle 1-q_i\rangle_i$. Consequently, the interquartile range—and therefore the estimated uncertainty—can become very small or even zero, which is an expected and reliable outcome under such conditions.

We note that IR excesses make disk-hosting sources easier to detect than their diskless counterparts, introducing a bias akin to the Malmquist bias and eventually leading to overestimated disk fractions in the lowest mass bin. To mitigate this, we restricted the computation of disk fractions to objects with estimated masses above the established sensitivity limits defined in \ref{sect:photometry}. Given the {\it unWISE}/W2 sensitivity, we estimate that our analysis is complete down to approximately 6\:\Mj\ when assuming 5\:Myr and about 8\:\Mj\ when assuming 10\:Myr, significantly extending the lower boundary of previous disk fraction studies in USC into the planetary-mass regime. Given these limits, no objects were excluded, as all members of the working sample had estimated masses above the thresholds under the two age assumptions.
Figure~\ref{fig:violin_fractions_woWbl} presents the resulting disk fractions, alongside the mass density distribution to illustrate how source filtering impacts different mass bins, most notably in the low-mass regime. This highlights the importance of filtering for potential contamination, especially where blending is more prevalent.

\begin{figure*}[h]
\begin{minipage}{0.5\linewidth}
    \centering
    \includegraphics[width=0.97\textwidth]{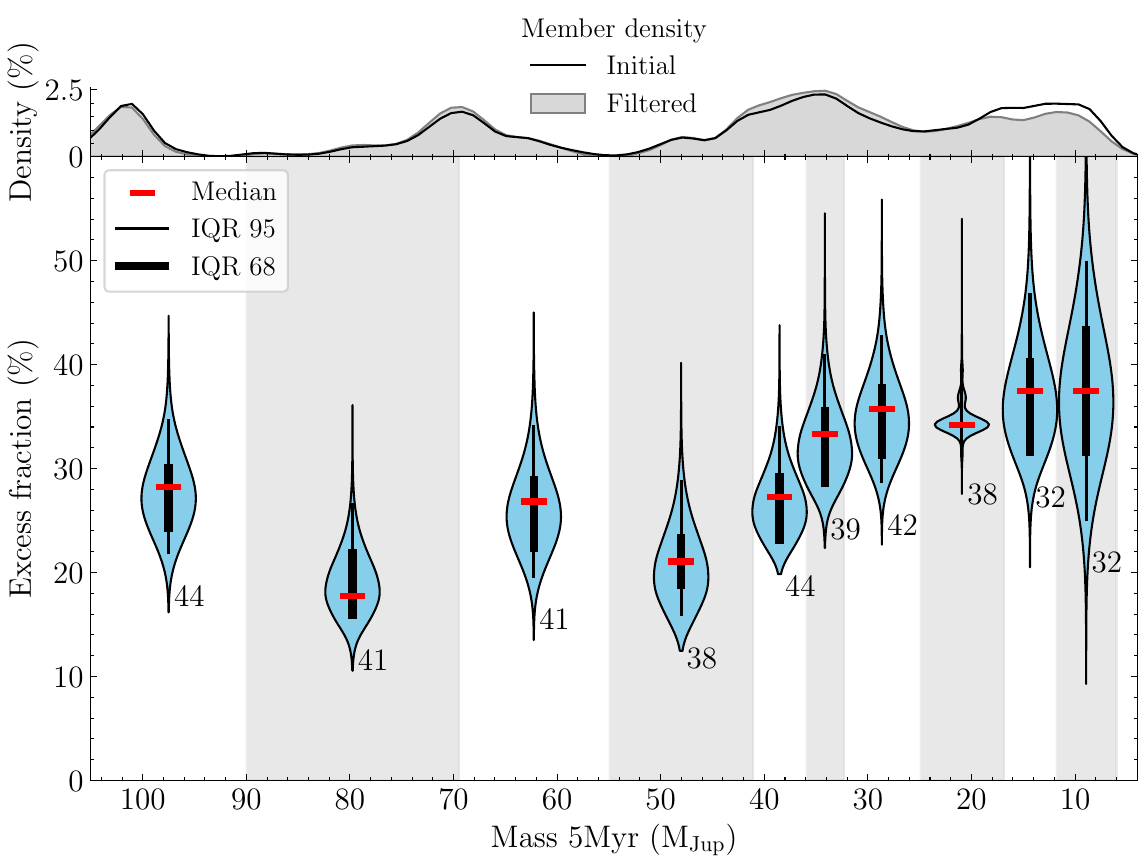}
\end{minipage}
\hfill
\begin{minipage}{0.5\linewidth}
    \centering
    \includegraphics[width=0.97\textwidth]{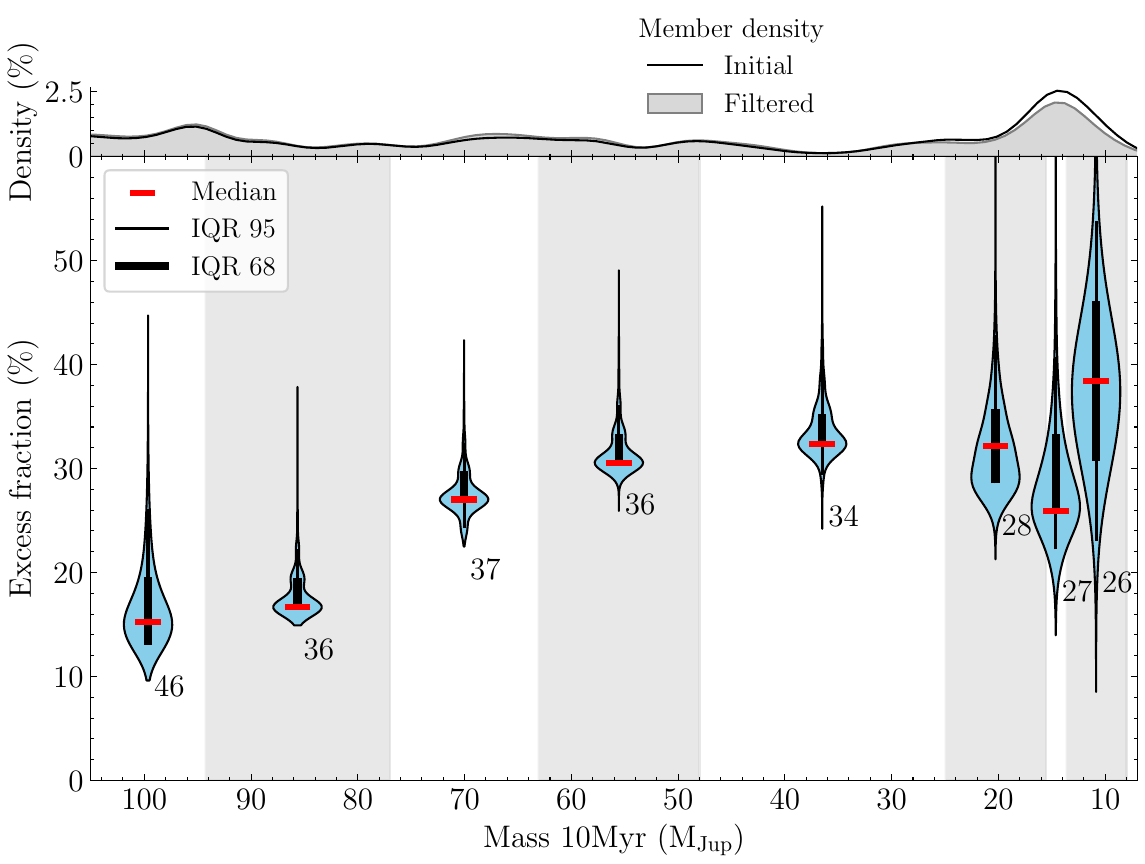}
\end{minipage}
\caption{Excess fraction as a function of central object mass for ages 5 ({\it Left}) and 10\:Myr ({\it Right}), over a violin graph of $\langle1-q_i^{(k)}\rangle_i$ distributions and mass density. Error bars were obtained from 95\% and 68\% (or 2 and 3$\sigma$ respectively) confidence intervals, calculated using the quantiles of the $\langle1-q_i^{(k)}\rangle_i$ distributions (see Sect.~\ref{sect:MCMC}). Red dots indicate the median of the distribution, adopted as the disk fraction. The numbers of objects included in each mass bin are indicated. Background shading is used to delineate mass bins. These distributions were obtained by removing the sources flagged in the {\it unWISE} images (see Sect.~\ref{sect:photometry}).}
\label{fig:violin_fractions_woWbl}
\end{figure*}

\section{Discussion}
\label{sect:discussion}

\subsection{A possible shift in trend near the substellar-planetary mass limit}
\label{sect:origin}

\begin{figure}[h]
\centering
\includegraphics[width=\columnwidth]{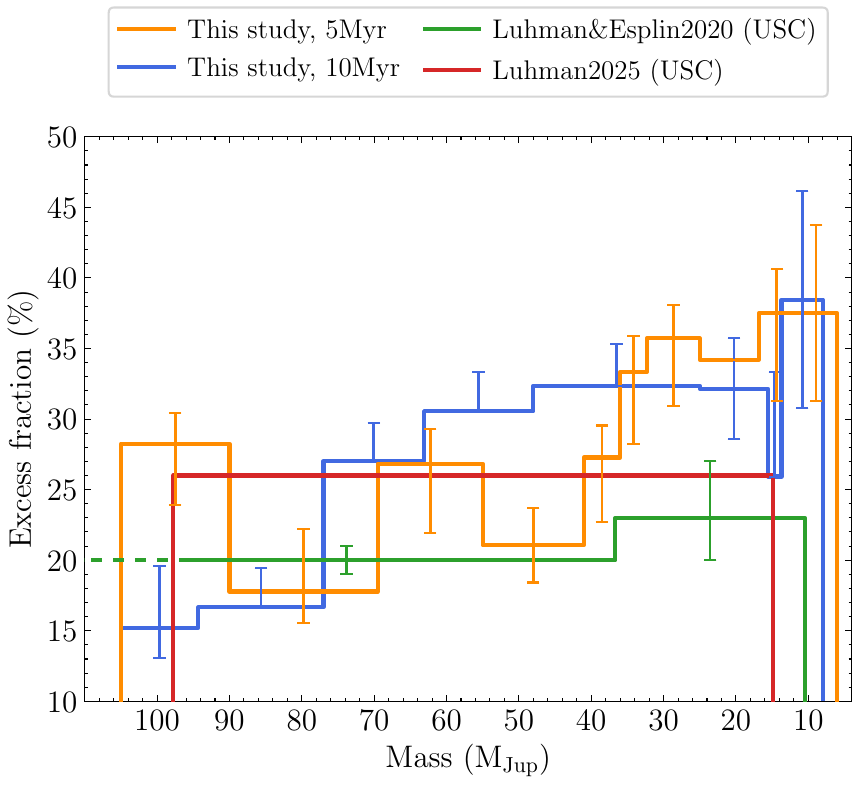}
\caption{Excess fractions as a function of central object mass in USC. Shown are the results from this study, for the masses inferred assuming ages of 5\:Myr and 10\:Myr. Error bars correspond to the 1$\sigma$ confidence interval derived from the $\langle1-q_i^{(k)}\rangle_i$ distribution (see Sect.~\ref{sect:result}). Literature values, summarized in Table~\ref{tab:fractions_lit}, are also shown.}
\label{fig:fractions_lit}
\end{figure}

Figure~\ref{fig:fractions_lit} provides a comparison between the disk fractions derived in this study with those reported by \cite{Luhman&Esplin2020} and \cite{Luhman2025} in USC. Overall, our measurements broadly align with previous estimates within the brown dwarf regime in USC \citep[approximately 20--25\%,][]{Luhman&Mamajek2012, Luhman&Esplin2020, Luhman2025, Dawson2013, Riaz2012}.
However, the use of broad mass bins in their analyses makes a quantitative comparison with our results difficult and may smooth out fine variations. Nevertheless, the figure reveals a continuous increase in disk fraction with decreasing mass, rising from the stellar into the substellar regime and reaching values of 30--35\% near the substellar-to-planetary mass boundary. This trend has been reported in previous studies \citep[e.g.,][]{Luhman2005c, Luhman2008b, Scholz&Jayawardhana2007, Ribas2015, Luhman&Esplin2020} and is commonly interpreted as evidence for mass-dependent disk lifetimes, which tend to be longer for lower-mass objects. While this general trend appears robust, its precise shape and amplitude likely depend on the specific physical conditions of each star-forming region. Indeed, some studies in other associations have reported disk fractions that show little to no mass dependence \citep{Scholz2023, Luhman2016}.

However, this trend seems to change in the lowest mass bins, with disk fraction flattening below 25\:\Mj\ for an assumed age of 5\:Myr or 45\:\Mj\ for an assumed age of 10\:Myr. The underlying real disk fraction at a given mass likely falls between these two borderline cases.
Figure~\ref{fig:fractions_lit} illustrates that the analyses by \citet{Luhman&Esplin2020} and \cite{Luhman2025}, which use only two and one broad mass bins in the substellar regime, respectively -- where we use between eight and ten bins (depending on the age) -- lack the resolution necessary to capture finer details, such as the flattening in disk fraction observed in our results.

However, this change in the slope should be interpreted with caution, as it could be the result of observational biases. As shown in Fig.~\ref{fig:MCMC_plot_woWbl} (and Appendix Fig.~\ref{fig:errors}), photometric uncertainties significantly increase toward lower masses. These larger uncertainties, as well as reduced mass coverage and the total number of points of the linear fit, contribute to a broader dispersion in the fitted parameters of the linear fits in the 0--25\:\Mj\ domain, as highlighted by the 3$\sigma$ shaded area in Fig.~\ref{fig:MCMC_plot_woWbl} and the associated uncertainties in Table~\ref{tab:MCMC_results_woWbl}. As a consequence, the boundary between excess-bearing and purely photospheric populations becomes more ambiguous in this mass regime. This effect is further amplified by the diminishing contrast of W2 excesses relative to the photospheric sequence at the lowest masses, as the peak of the photospheric emission shifts to longer wavelengths. This trend has been previously noted in the analogous {\it Spitzer}/IRAC2 band (4.5\:$\mu$m) by \citet{Luhman2010, Luhman&Mamajek2012, Luhman2006}.

For these reasons, the statistical power of both the linear regression and the identification of excesses (constrained by increasing photometric uncertainties) is diminished in this regime. Consequently, the estimated excess fraction in the planetary-mass domain may underestimate the true underlying disk frequency. The apparent change in slope near the substellar-to-planetary mass boundary should therefore be interpreted with caution. Longer-wavelength excesses such as in W3 and W4 would have been stronger and better suited to overcome this attenuation but remain unusable in our case, as discussed in Sect.~\ref{sect:photometry}.

\subsection{Implications for the formation process of FFPs and low-mass BDs}
\label{sect:implications}

If confirmed, the observed shift in trend near the substellar-planetary mass limit may have important implications for the origin of these objects and could suggest a shift in the relative contribution of different formation pathways. Interestingly, the mass at which this flattening occurs—below 25\:\Mj\ at 5\:Myr—is consistent with the change in slope in the IMF reported by \cite{MiretRoig2022a}, indicating an excess of objects compared to core-collapse predictions and suggesting that dynamical ejection from planetary systems may play a significant role in the formation of FFPs.

Ejection processes are known to truncate circumstellar disks or, in some cases, strip them entirely \citep{Umbreit2011, Testi2016, Rabago&Steffen2019}, which naturally leads to a reduced disk fraction among ejected BDs and FFPs. If real, the observed change in the disk fraction trend near 25\:\Mj\ could therefore reflect an increasing prevalence of ejection events below this mass threshold. Interestingly, \citet{Scholz2023} reported that only one out of six NGC\:1333 members with spectral types of L0 or later shows evidence of a disk, echoing the trend observed in our own sample. They suggest that, if this drop-off at the lowest masses is confirmed, it may indicate that FFPs are predominantly formed through ejection rather than star-like formation processes. However, they also caution that the small sample size and associated statistical uncertainties limit the significance of this result.

However, this interpretation is complicated by the findings of \citet{Umbreit2011}, whose simulations of stellar and planetary encounters suggest that while such interactions do truncate disks, they can also significantly enhance the surface density in the innermost disk regions. Since our (W1–W2) based analysis is most sensitive to emission from these inner regions, this densification could, in principle, increase the detectability of disks following ejection. This effect, however, may be short-lived if the enhanced inner disk density is rapidly dissipated through viscous evolution or increased accretion onto the central object, rendering the excess emission transient. Hence, its timescale needs to be further investigated.

Additionally, simulations by \citet{Parker&AlvesdeOliveira2023} suggest that FFPs formed via ejection may exhibit higher mean velocities, causing them to rapidly migrate away from the core of their parent associations. As a result, such objects would likely be underrepresented in both our study and previous disk frequency surveys, potentially introducing a bias against the lowest-mass population.

\subsection{Comparison with the literature and disk evolution}
\label{sect:comparison}

\renewcommand{\arraystretch}{1.2}
\begin{table*}[t]
\centering
\caption{Disk fractions from this work and the literature in various regions.}
\label{tab:fractions_lit}
\begin{tabular}{ll r@{\,--\,}l r@{\,--\,}l cccc}
\toprule
References &
Disk fraction &
\multicolumn{2}{c}{{SpT range}} &
\multicolumn{2}{c}{{Mass range$^a$}} &
Region &
Age$^b$ &
$N_\mathrm{objects}$$^d$ &
Method \\
&
(\%) &
\multicolumn{2}{c}{} &
\multicolumn{2}{c}{{(\Mj)}} &
&
(Myr) &
& \\
\midrule
This work & $37.5\pm6.2$         & & & 6    & 11.8 & Upper Sco & 5 & 32 & W1--W2 vs Mass \\
          & $37.5_{-6.2}^{+3.1}$ & & & 11.8 & 16.8 &           & 5 & 32 &                \\
          & $34.2\pm0.0$         & & & 16.8 & 25   &           & 5 & 38 &                \\
 
          & $38.5_{-7.7}^{+7.7}$ & & & 8    & 13.7 &           & 10 & 24 &               \\
          & $25.9_{-0.0}^{+7.4}$ & & & 13.7 & 15.5 &           & 10 & 27 &               \\
          & $32.1\pm3.6$         & & & 15.5 & 25   &           & 10 & 28 &               \\
\midrule
\citet{Luhman\string&Esplin2020} & $23_{-3}^{+4}$ & M8 & L2 & 10.5 & 36.7 & Upper Sco & 10 & 27/115 & K$_{\rm s}$--I2/W2 vs SpT \\
\citet{Luhman2025} & $26$ & M6.25 & M9.5 & 14.8 & 97.8 & Upper Sco & 10 & 52/200 & W1--W2 vs SpT \\
\citet{Luhman2016} & $52\pm10$ & M8 & L0 & 10.6 & 23 & NGC\:1333 & 3 & 11/21 & Literature ({\it Spitzer}) \\
\citet{Scholz2023} & $42_{-16}^{+18}$ & M9 & L3 & 5.9 & 12.0 & NGC\:1333 & 2 & 5/12 & K--I2 vs SpT \\
\citet{Esplin\string&Luhman2019} & $41_{-7}^{+8}$ & M8 & L0 & 14.2 & 23.7 & Taurus & 3 & 15/37 & K$_{\rm s}$--I2/W2 vs SpT \\
\citet{Damian2023} & $80\pm55$ $^c$ & & & 10.5 & 32.5 & $\sigma$\:Orionis & 2\,--\,3 & 5/6 & K$_{\rm s}$--W2 slope \\
%\citet{PenaRamirez2012} & $>32\pm11$ $^d$ & M9\:--\:T1 & 4.2\:--\:12.6 & $\sigma$\:Orionis & 2--3 & -- & J--I2 vs Z--J \\
\citet{Luhman2016} & $42_{-9}^{+11}$ & M8 & L0 & 10.6 & 23 & IC\:348 & 3 & 10/24 & Literature ({\it Spitzer}) \\
\citet{Seo\string&Scholz2025} & $46_{-12}^{+13}$ & M9 & L1 & 9.3 & 13.2 & IC\:348 & 3 & 6/13 & K--IRAC2 vs SpT \\
& & \multicolumn{2}{c}{} & \multicolumn{2}{c}{} & & & & + SED modeling \\
\bottomrule
\end{tabular}
\tablefoot{
$^a$ Mass range when not explicitly given in the referenced study, inferred using the \citet{Faherty2016} temperature scale for young brown dwarfs and the evolutionary models of \citet{Saumon2008}, based on the spectral type and the age$^b$ of the region. \\
$^b$ Age of the region as adopted by the authors. Note that these values may differ across studies and may have been revised in light of recent constraints, particularly following the release of Gaia data \citep[e.g., Gaia DR2 and DR3;][]{GaiaDR2, GaiaDR3}. \\
$^c$ Not shown in Fig.~\ref{fig:fractions_lit} to preserve clarity due to its extreme value and ill-defined error. \\
%$^d$ reported as a lower limit; 
$^d$ Number of objects in each mass bin for this study. Fractions expressed as N$_\mathrm{disk}$/N$_\mathrm{tot}$ represent those found in the literature values, as reported by the authors. \\
Fields marked with “--” indicate values not specified by the authors.}
\end{table*}

\begin{figure}[h]
\centering
\includegraphics[width=\columnwidth]{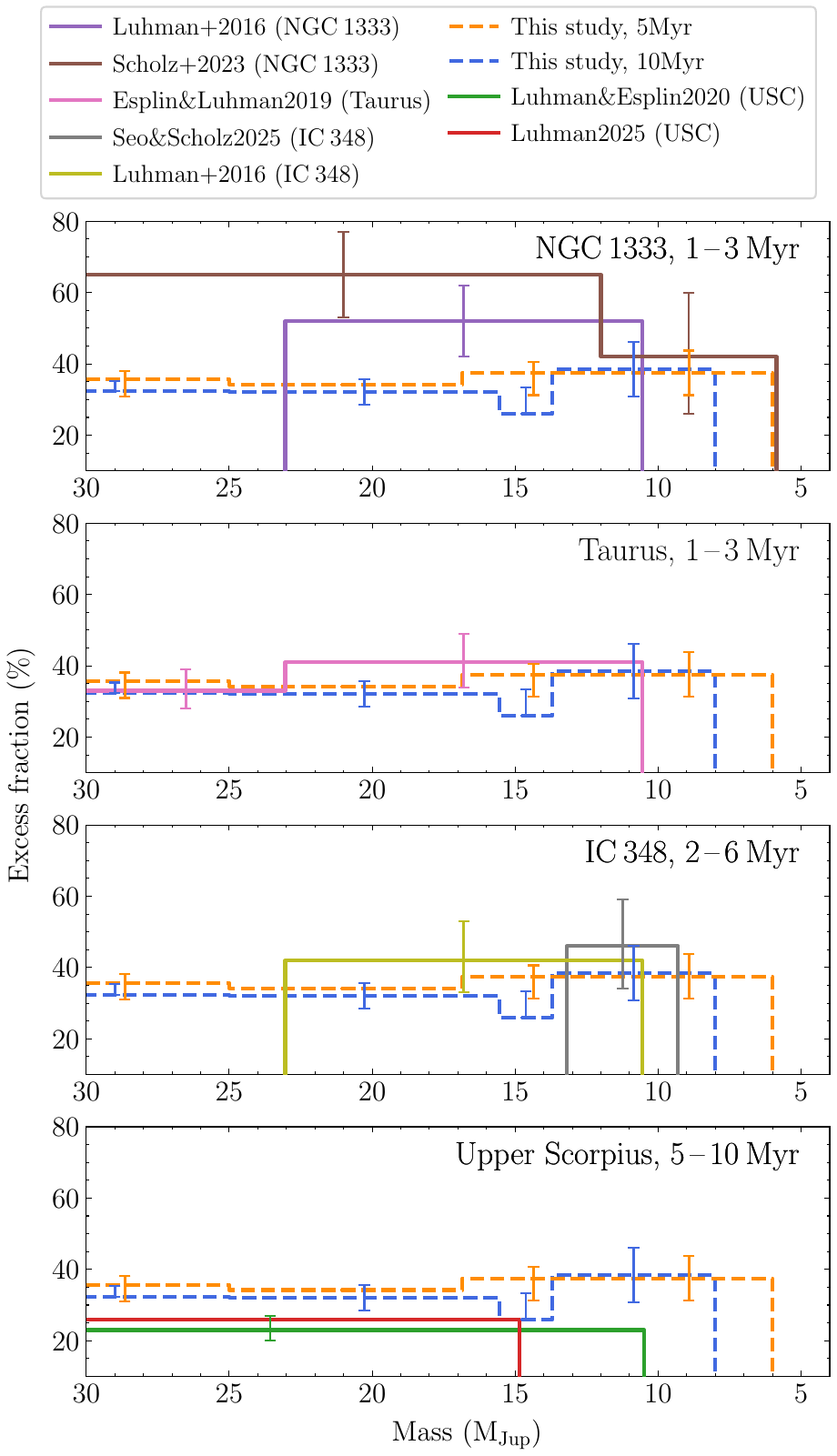}
\caption{Excess fractions as a function of central object mass below 30\:\Mj. Shown are the results from this study for masses inferred assuming ages of 5\:Myr and 10\:Myr. Error bars correspond to the 1$\sigma$ confidence interval derived from the $\langle1-q_i^{(k)}\rangle_i$ distribution (see Sect.~\ref{sect:result}. Literature values, summarized in Table~\ref{tab:fractions_lit}, are also shown.}
\label{fig:fractions_lit_zoom}
\end{figure}

Table~\ref{tab:fractions_lit} gives an overview of the disk fractions among BDs and FFPs reported in the literature in various associations and clusters over the past few years. All these studies derived disk fractions based on IR excesses detected in either the {\it WISE}/W2 (4.6\:$\mu$m) or {\it Spitzer}/I2 (4.5\:$\mu$m) bands. Thus, they probe the same component of primordial disks and employ methodologies broadly comparable to ours, enabling meaningful and consistent comparisons of disk fraction estimates. We note, however, that the uncertainties reported in these studies were typically derived using standard binomial approximations, which are statistically less robust than the probabilistic framework adopted in our analysis (see Sect.~\ref{sect:result}) and tend to underestimate the true uncertainties. 

Figure~\ref{fig:fractions_lit_zoom} presents the disk fractions derived in this work, below 30\:\Mj\ ,  along with the values summarized in Table~\ref{tab:fractions_lit}. We omit the single point from \citet{Damian2023} due to its very large uncertainties.
Although Fig.~\ref{fig:fractions_lit_zoom} offers a comparative view of disk fractions among FFPs across different studies, interpreting these results remains inherently difficult. First, the small sample sizes used in these studies led to large statistical uncertainties and broad mass bins. Their heterogeneous mass coverage led to a heterogeneous binning of the mass domain. In this context, the present analysis provides the largest (and hence most statistically robust) sample, allowing for a much finer sampling of the substellar and planetary-mass domain. In spite of these limitations, a number of trends can be seen in Fig.~\ref{fig:fractions_lit_zoom}. 

First, the disk fractions reported in younger regions -- namely NGC\:1333 (1–3\:Myr) by \citet{Scholz2023}, Taurus (1–3\:Myr) by \citet{Esplin&Luhman2019}, and IC\:348 (2–6\:Myr) by \citet{Seo&Scholz2025} -- are slightly higher but are consistent within the uncertainties, reaching 40--50\%, compared to the $\sim$35\% we measure in USC. If this decrease with age is real, it might reflect intrinsic disk dispersal timescales, consistent with USC’s older age relative to these associations and the well-established decline in disk fraction over time. Interestingly, the disk fraction among 15--25\:\Mj\ objects in Taurus appears intermediate between USC and NGC\:1333, possibly reflecting the younger average age of the latter.

Second, this comparison underscores that, even at the age of USC, the disk fractions in the substellar and planetary-mass regime remain substantial -- less than a quarter lower than those observed in younger regions (e.g., NGC\:1333, Taurus, and IC\:348). This moderate decline over 3$\sim$5\:Myr suggests relatively extended dispersal timescales for disks around the lowest-mass objects, with significant implications for the potential formation of companions and moons, and offers interesting prospects for exploring hierarchical isolated planetary systems.

Current interpretations remain largely limited by the scarcity and large uncertainties of disk fraction estimates near and within the planetary-mass regime. To confidently trace trends -- including the possible flattening in disk fractions we observe -- and to enable meaningful comparisons across various regions, larger samples in the aforementioned regions are needed to improve bin statistics and enhance mass resolution. The unprecedented sensitivity and spatial resolution of JWST will be crucial in overcoming these limitations and providing key insights into the formation pathways and the environmental dependence of FFP disk evolution. Furthermore, capturing the elusive debris and more evolved disks around FFPs would be highly valuable for shedding light on the transition from primordial to debris disks, carrying significant implications for disk lifetimes and evolution mechanisms.

\section{Conclusion}
\label{sect:conclusion}

In this study, we investigated the disk fraction among the USC members reported in \cite{MiretRoig2022a} in the mass range between a few Jupiter masses and 105\:\Mj. Depending on the assumed age, our working sample includes between 17 and 40 objects below 13\:\Mj, offering a sample size that is 1.4 to 3.3 times larger than the most comprehensive studies to date by \cite{Scholz2023} and \cite{Seo&Scholz2025}.

We used {\it unWISE} mid-IR photometry to identify IR excesses in the (W1--W2) color associated with the presence of primordial disks and computed the disk fraction as a function of the estimated masses reported in \cite{MiretRoig2022a}. We estimate that the {\it unWISE} images are complete down to $\sim$6\:\Mj\ assuming 5\:Myr, or $\sim$8\:\Mj\ assuming 10\:Myr, substantially extending the lower boundary of previous disk fraction studies into the planetary-mass regime. We carried out a detailed visual inspection of the {\it unWISE} images and flagged potentially unreliable observations. 

We used a Bayesian outlier detection framework to robustly model the empirical photospheric sequence of diskless sources in the (W1–W2) color and to identify objects exhibiting IR excesses. This approach explicitly incorporates outlier modeling, allowing us to compute the probability that each source harbors a disk-related IR excess. These individual probabilities enabled us to derive posterior distributions of excess fractions across finely sampled mass bins extending with unprecedented resolution into the planetary-mass regime, along with statistically sound estimates of the associated uncertainties.

To assess the reliability of our method, we compared our excess detections to those reported by \citet{Luhman&Esplin2020} for a set of objects in common with known spectral types. The excellent agreement between the two approaches, with discrepancies limited to under 5\%, lends robust support to the validity of our methodology.

We confirm that the disk fraction continuously increases with decreasing mass, from the stellar (105\:\Mj) into the substellar regime, exceeding 30\% near the substellar-to-planetary mass boundary, a trend commonly observed and attributed to longer disk lifetimes for lower-mass objects. We identify a possible flattening in this trend near the substellar-to-planetary mass boundary ($\sim$25--45\:\Mj\ , assuming ages of 5--10\:Myr). If confirmed, this plateau may reflect a shift in the dominant formation mechanism, such as a greater prevalence of dynamical ejection, which is expected to truncate or strip disks. Alternatively, the apparent change could arise from observational limitations. In particular, increasing photometric uncertainties and a redder photospheric (W1--W2) color at the lowest masses tend to blur the distinction between disk-bearing and diskless populations, reducing the sensitivity of excess detection and potentially biasing disk fraction estimates downward.

Comparisons with values from the literature suggest that disk fractions in USC are roughly a quarter lower than those observed in younger regions (e.g., Taurus, NGC\:1333, and IC\:348) within and near the planetary-mass regime. This difference is broadly consistent with expected disk dispersal over time. Overall, placing these results in the context of previous studies remains challenging due to the small sizes of previous samples (typically less than a dozen objects), which leads to large uncertainties and coarse mass binning. This underscores the pressing need for large, homogeneous samples to enable meaningful inter-region comparisons.

While not pointing to a single dominant formation scenario, our analysis places new constraints on the disk fraction down to the planetary-mass regime with a fine sampling of the mass domain between 6--105\:\Mj, setting the stage for further systematic investigations. Additional follow-up, particularly with JWST, should confirm or rule out the shift in trend seen around $\sim$25--45\:\Mj. Similarly, ALMA follow-up observations will be essential to explore the properties of these disks -- particularly their masses and eventually their sizes -- down to the planetary-mass regime, and to potentially distinguish between disks formed via core collapse and those stripped of material through dynamical ejection.

\section*{Data availability}

Tables~\ref{tab:members_sample} is only available in electronic form at the CDS via anonymous ftp to \texttt{cdsarc.u-strasbg.fr} (130.79.128.5) or via \url{http://cdsweb.u-strasbg.fr/cgi-bin/qcat?J/A+A/}.

\begin{acknowledgements}
We thank our anonymous referee for their timely and constructive report, which has helped improve this manuscript.
We thank Gaspard Duchêne for the helpful discussions.
This research has received funding from the European Research Council (ERC) under the European Union’s Horizon 2020 research and innovation programme (grant agreement No 682903, P.I. H. Bouy), and from the French State in the framework of the ”Investments for the future” Program, IdEx Bordeaux, reference ANR-10-IDEX-03-02.
This study received financial support from the French government in the framework of the University of Bordeaux's France 2030 program / RRI "ORIGINS".
D.B. and N.H. are supported by grants No.PID2019-107061GB-C61 and PID2023-150468NB-I00 by the Spanish Ministry of Science and Innovation/State Agency of Research MCIN/AEI/10.13039/501100011033.
M.T. is supported by JSPS KAKENHI grant No.24H00242.
P.A.B. Galli acknowledges financial support from São Paulo Research Foundation (FAPESP) under grant 2020/12518-8.
This publication makes use of VOSA, developed under the Spanish Virtual Observatory (https://svo.cab.inta-csic.es) project funded by MCIN/AEI/10.13039/501100011033/ through grant PID2020-112949GB-I00. VOSA has been partially updated by using funding from the European Union's Horizon 2020 Research and Innovation Programme, under Grant Agreement nº 776403 (EXOPLANETS-A).
This research has made use of the VizieR catalogue access tool, CDS, Strasbourg, France. The original description of the VizieR service was published in A\&AS 143, 23.
This work made use of GNU Parallel \citep{Tange2011a}, astropy \citep{astropy2013, astropy2018}, Topcat \citep{Topcat}, matplotlib \citep{matplotlib}, Numpy \citep{numpy}, APLpy \citep{aplpy2019}, dustmaps \citep{dustmaps}, SPLAT \citep{Burgasser2017}, astroML \citep{astroML, astroMLText}, PyMC \citep{pymc}.
\end{acknowledgements}

\bibliography{biblio}

\begin{appendix}

\renewcommand{\topfraction}{0.95}
\renewcommand{\textfraction}{0.05}
\renewcommand{\floatpagefraction}{0.9}

\FloatBarrier
\section{Complementary analysis of the entire sample including flagged sources}
\label{app:analysis_including_flagged}

\begin{figure*}[b!]
\centering
\includegraphics[width=0.931\textwidth]{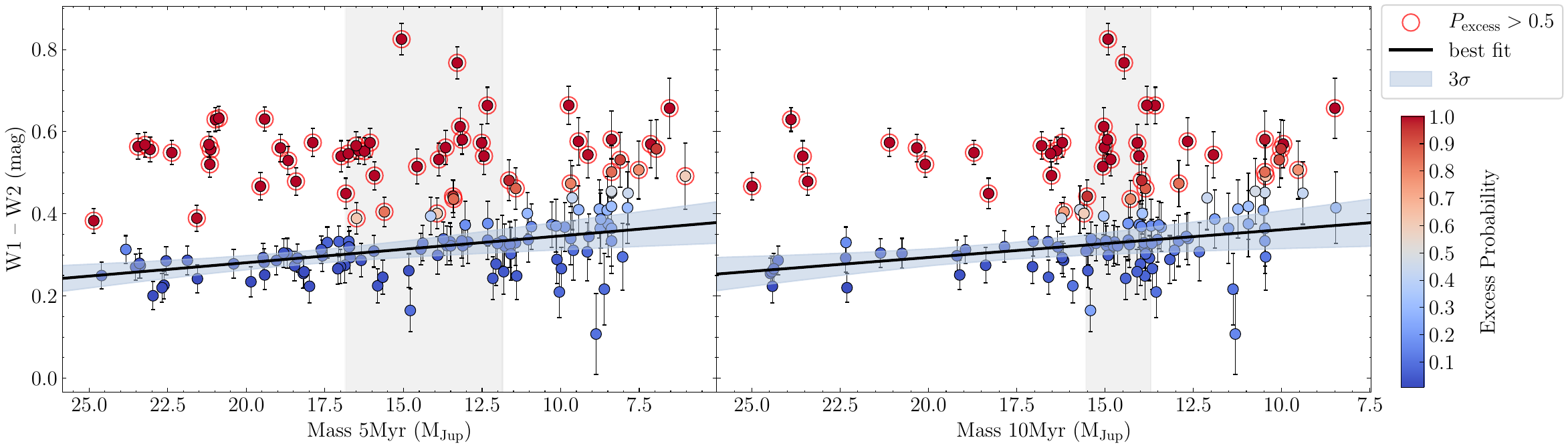}
\caption{W1--W2 IR color as a function of mass for USC members in the 0--25\:\Mj\ mass range. The data are illustrated for assumed ages of 5 and 10\:Myr and for the entire sample keeping flagged sources in {\it unWISE} images (see Sect.~\ref{sect:photometry}). This zoom-in focuses on the low-mass regime to better visualize the linear fits and the associated excess probabilities in this domain, as shown in Fig.~\ref{fig:MCMC_plot} and discussed in Sect.~\ref{sect:MCMC}. See the caption in Fig.~\ref{fig:MCMC_plot} for a description.}
\label{fig:MCMC_plot_lows}
\end{figure*}

\begin{figure}[H]
\centering
    \includegraphics[width=\columnwidth]{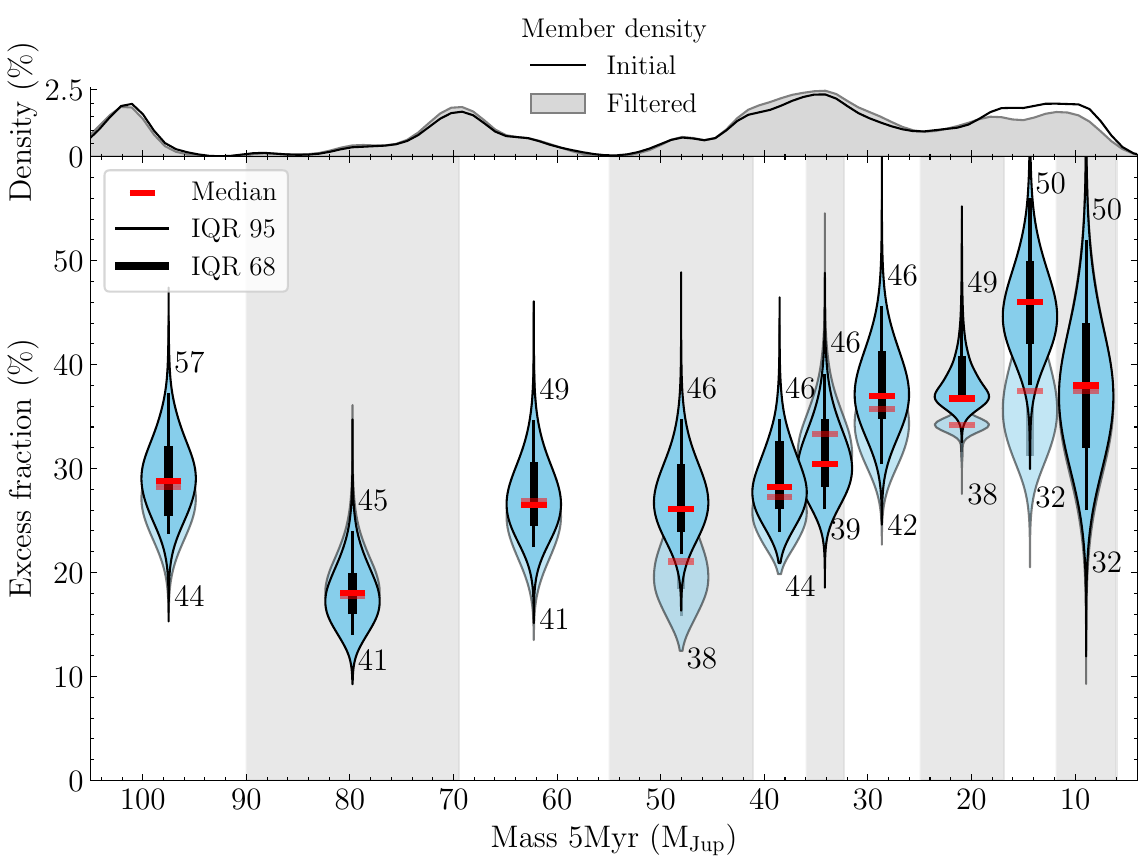}
    \includegraphics[width=\columnwidth]{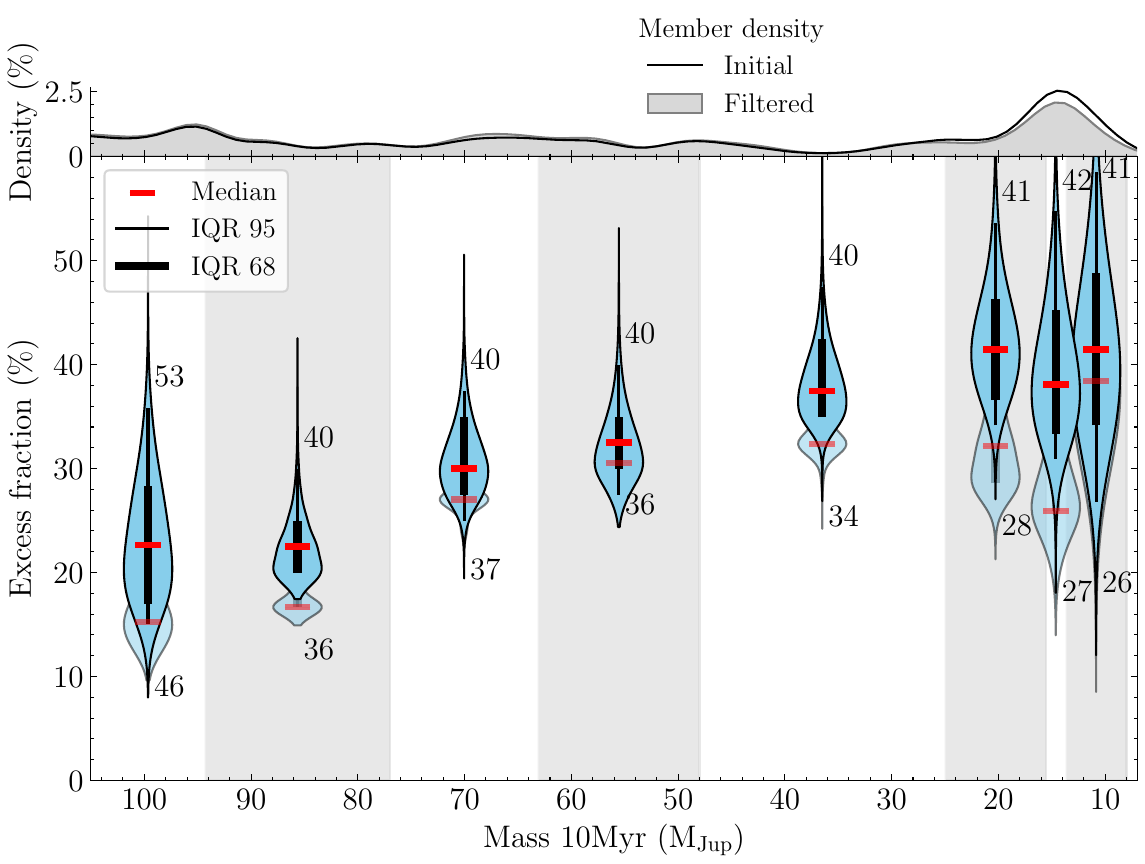}
\caption{Excess fraction as a function of central object mass for assumed ages of 5 ({\it Top}) and 10\:Myr ({\it Bottom}), over a violin graph of the $\langle1-q_i^{(k)}\rangle_i$ distributions and mass density. Same features as Fig.~\ref{fig:violin_fractions_woWbl} are shown. Linear regression was performed for the entire sample keeping flagged sources in {\it unWISE} images (see Sect.~\ref{sect:photometry}). The original fractions for the cleaned sample are shown in the background with transparency for comparison.}
\label{fig:violin_fractions}
\end{figure}

\begin{figure*}[htpb]
\centering
\includegraphics[width=\textwidth]{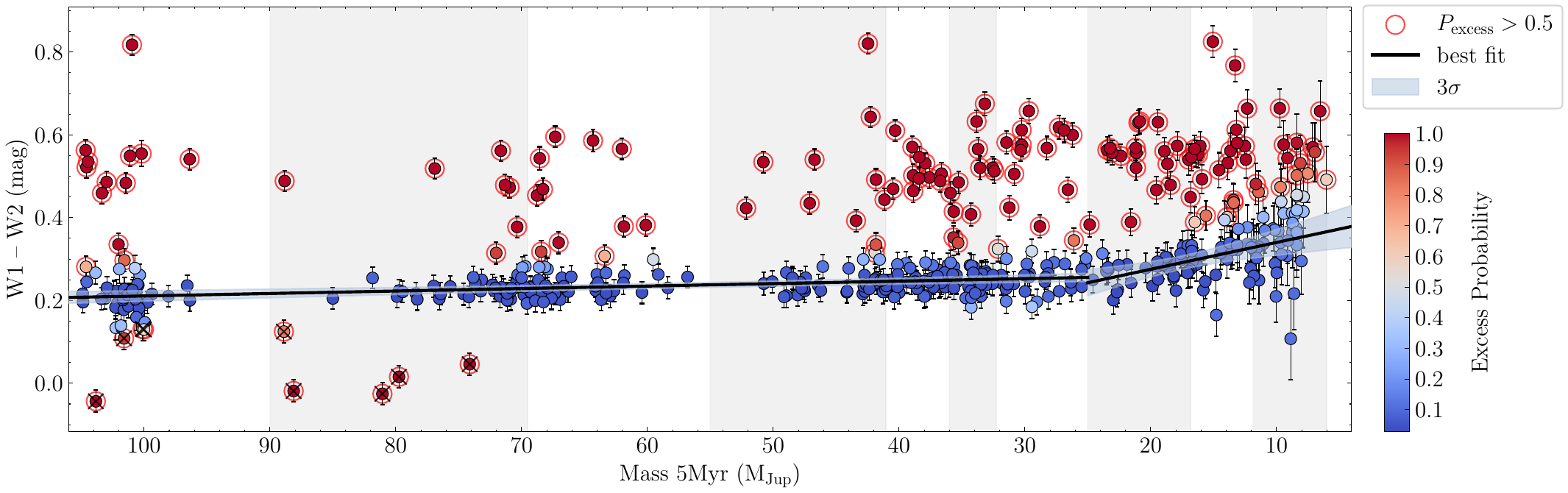}
\includegraphics[width=\textwidth]{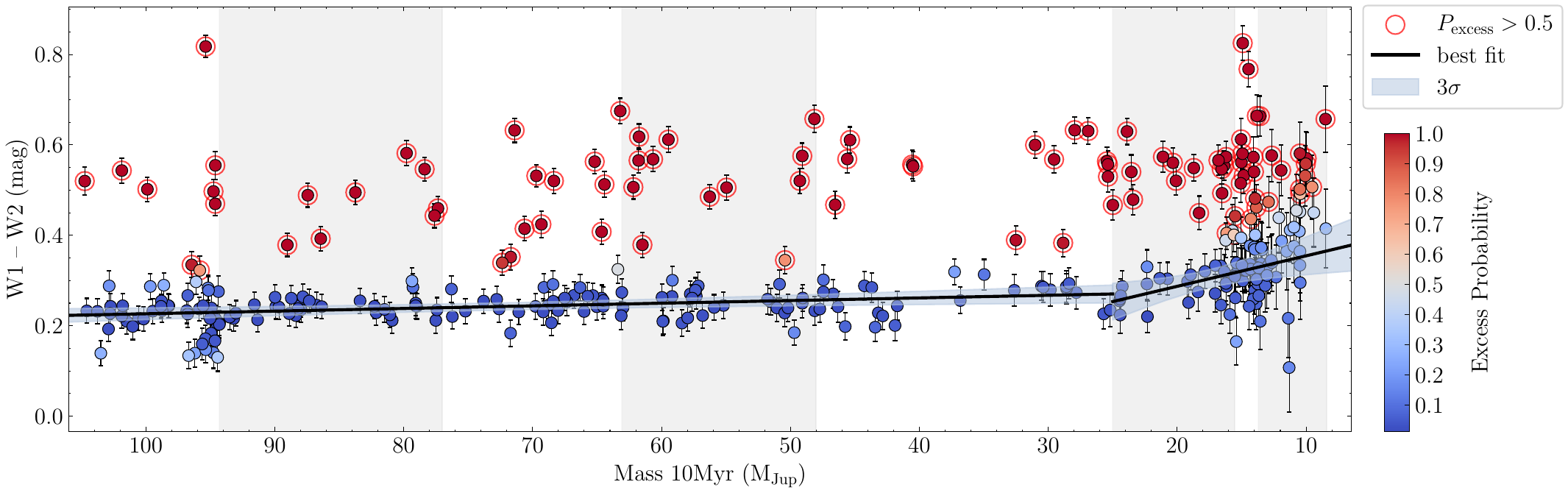}
\caption{W1--W2 IR color as a function of mass for USC members, assuming ages of 5 ({\it Top}) and 10\:Myr ({\it Bottom}), for the entire sample keeping flagged sources in {\it unWISE} images (see Sect.~\ref{sect:photometry}). The solid black line represents the best-fit relation with outlier rejection applied, while the blue-shaded region corresponds to the 3$\sigma$ confidence interval of the fit parameters (i.e., slope and intercept). The color scale represents the inferred excess probability, as indicated by the color bar. Red circles highlight sources with an excess probability greater than 0.5. Objects below the fitted photospheric sequence identified as outliers ($P_{\mathrm{excess}}>0.5$) are marked with black crosses, since they are excluded from the subsequent analysis. Background shading is used to delineate mass bins used in Figure~\ref{fig:violin_fractions}.}
\label{fig:MCMC_plot}
\end{figure*}

For completeness, and acknowledging that not all conservatively flagged sources necessarily suffer from contamination in the W1 and W2 bands, we also performed the analysis for the whole sample, keeping these flagged sources. The corresponding results are shown in Figures~\ref{fig:MCMC_plot} and \ref{fig:MCMC_plot_lows}, with derived fit parameters summarized in Table~\ref{tab:MCMC_results}. 
The resulting disk fractions are shown in Fig.~\ref{fig:violin_fractions}, with those for the clean sample overlaid in transparency for comparison, and are listed in Table~\ref{tab:fractions}. 
When the comparison with \cite{Luhman&Esplin2020} is extended to include flagged sources, the discrepancy rates increase only slightly to 7\% and 8\%, respectively. 
This difference between the flagged and full sample is likely the result of spurious excess detections, as discussed in Sect.~\ref{sect:photometry}.
Notably, the majority of flagged sources in our analysis exhibit IR excesses, and their addition leads to a significant increase in the disk fraction at the lowest masses, depending on the specific mass bin (see Fig.~\ref{fig:violin_fractions}). This suggest that contamination from unresolved companions may account for a non-negligible fraction of excess detections among low-mass BDs and FFPs. If contamination were not a contributing factor, we would expect the flagged sources to be more evenly distributed between excess-bearing and purely photospheric objects.

\begin{figure}[!h]
    \centering
    \begin{minipage}{0.49\linewidth}
        \centering
        \includegraphics[width=\linewidth]{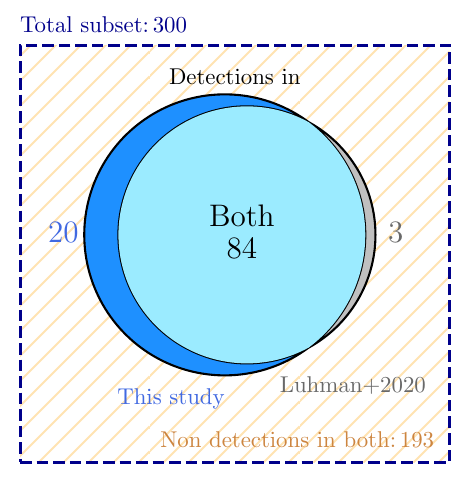}
    \end{minipage}
    \hfill
    \begin{minipage}{0.49\linewidth}
        \centering
        \includegraphics[width=\linewidth]{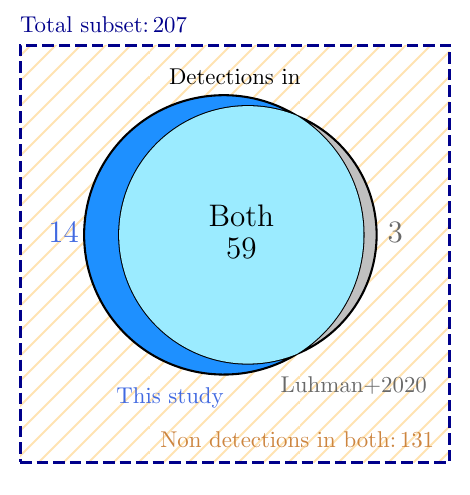}
    \end{minipage}
    \caption{Comparison of W2 excess detections between this study and \cite{Luhman&Esplin2020} using Venn diagrams. The data are illustrated for assumed ages of 5 ({\it Left}) and 10\:Myr ({\it Right}), for the entire sample keeping flagged sources in {\it unWISE} images (Sect.~\ref{sect:photometry}). In each case, the central intersection shows the number of sources identified as exhibiting excess in both studies, while the left and right circle arcs show those detected only in one study but not the other. The orange-hatched region represents sources without excess detection in either work.}
    \label{fig:venn}
\end{figure}

\FloatBarrier

\renewcommand{\arraystretch}{1.4}
\begin{table*}
\centering
\caption{Posterior estimates of the key regression and outlier model parameters from our Bayesian inference for the entire sample.}
\label{tab:MCMC_results}
\begin{tabular}{lcccc}
\toprule
\multirow{2}{*}{Parameters} & \multicolumn{2}{c}{\textbf{Mass 5 Myr}} & \multicolumn{2}{c}{\textbf{Mass 10 Myr}} \\
\cmidrule(lr){2-3} \cmidrule(lr){4-5}
                            & \hspace{2mm}$[25$--$105]$\:\Mj\hspace{3mm} & \hspace{3mm}$[0$--$25]$\:\Mj \hspace{3mm} 
                            & \hspace{3mm}$[25$--$105]$\:\Mj \hspace{3mm} & \hspace{3mm}$[0$--$25]$\:\Mj \vspace{1mm} \\ 
\midrule
$\text{intercept}$ & $0.271_{-0.010}^{+0.010}$     & $0.405_{-0.039}^{+0.042}$   & $0.285_{-0.018}^{+0.017}$     & $0.422_{-0.057}^{+0.053}$ \\
$\text{slope}$     & $-0.0006_{-0.0002}^{-0.0002}$ & $-0.0065_{-0.0023}^{+0.0022}$ & $-0.0006_{-0.0002}^{+0.0002}$ & $-0.0068_{-0.0030}^{+0.0030}$ \\
\addlinespace
$P_b$              & $0.305_{-0.058}^{+0.053}$     & $0.406_{-0.095}^{+0.100}$   & $0.292_{-0.076}^{+0.078}$     & $0.411_{-0.120}^{+0.129}$ \\
$Y_b$              & $0.419_{-0.037}^{+0.037}$     & $0.525_{-0.037}^{+0.035}$   & $0.480_{-0.052}^{+0.049}$     & $0.516_{-0.051}^{+0.046}$ \\
$\sigma_b$         & $0.172_{-0.022}^{+0.026}$     & $0.090_{-0.023}^{+0.033}$   & $0.125_{-0.034}^{+0.045}$     & $0.098_{-0.031}^{+0.042}$ \\
\bottomrule
\end{tabular}
\tablefoot{The entire sample keeps flagged sources in {\it unWISE} images (see Sect.~\ref{sect:photometry}). We separately show results for the high-mass ($25-105$\:\Mj) and low-mass ($0-25$\:\Mj) regimes and for the analysis with masses assuming ages of 5 and 10\:Myr. The mass division between the low- and high-mass regimes is set at $M = 25$\:\Mj, following the observed break in the sequence (see Sect.~\ref{sect:MCMC}). Reported values correspond to the posterior median, with uncertainties given as 3--97\% HDIs in subscript and superscript.
$P_b$ is the global outlier fraction in the population. $Y_b$ and $\sigma_b$ are the center and scale of the outlier (excess) distribution in W1--W2 color space. The slope and intercept correspond to the best-fit linear regression for the photospheric sequence. Posteriors are well-converged with $\hat{R} \simeq 1.0$ for all parameters.}
\end{table*}

\renewcommand{\arraystretch}{1.1}
\begin{landscape}
\vspace*{\fill}
\begin{table}[H]
\caption{Sample of members analyzed in this study.}
\label{tab:members_sample}
\centering

\begin{tabular}{%lccccccccccccccc}
    l       % ID
    T{3.5}  % RA
    T{2.5}  % Dec
    T{3.2}  % d
    T{1.2}  % sigma_d
    T{3.2}  % Mass 5 Myr
    T{3.2}  % Mass 10 Myr
    T{2.3}  % W1
    T{1.4}  % sigma_W1
    T{2.3}  % W2
    T{1.4}  % sigma_W2
    c                    % flag
    T{1.3}  % p_exc 5 Myr
    T{1.3}  % p_exc 10 Myr
    T{1.3}  % p_exc 5 Myr w
    T{1.3}  % p_exc 10 Myr w
}
\toprule
ID & 
{RA (J2000)} & 
{Dec (J2000)} & 
{$d$} & 
{$\sigma_d$} & 
{Mass 5\:Myr} & 
{Mass 10\:Myr} & 
{W1} & 
{$\sigma_\mathrm{W1}$} & 
{W2} & 
{$\sigma_\mathrm{W2}$} & 
{flag} & 
{$p_\mathrm{exc}^\mathrm{\,5\,Myr}$} & 
{$p_\mathrm{exc}^\mathrm{\,10\,Myr}$} & 
{$p_\mathrm{exc}^\mathrm{\,5\,Myr,\,\mathrm{w}}$} & 
{$p_\mathrm{exc}^\mathrm{\,10\,Myr,\,\mathrm{w}}$} \\
DANCe & {(deg)} & {(deg)} & {(pc)} & {(pc)} & {(\Mj)} & {(\Mj)} & {(mag)} & {(mag)} & {(mag)} & {(mag)} & & & & & \\
\midrule
126 & 242.56423	& -29.39899 & 130.09 & 1.40 & 68.68  & 137.27 & 11.884 & 0.0178 & 11.640 & 0.0195 & 0 & 0.043 & {--}  & 0.042 & {--}   \\
173 & 241.73788	& -27.72747 & 153.16 & 3.63 & 33.46  & 66.80  & 12.522 & 0.0190 & 12.255 & 0.0223 & 0 & 0.055 & 0.034 & 0.050 & 0.011  \\
180 & 243.58587	& -27.76383 & 155.35 & 6.09 & 29.63  & 59.19  & 13.183 & 0.0195 & 12.918 & 0.0238 & 0 & 0.053 & 0.028 & 0.051 & 0.010  \\
181 & 243.05388	& -27.88690 & 157.25 & 3.54 & 33.44  & 67.33  & 12.933 & 0.0191 & 12.680 & 0.0230 & 0 & 0.043 & 0.024 & 0.039 & 0.007  \\
201 & 242.47569	& -27.41861 & 108.34 & 0.63 & 101.14 & 165.10 & 11.477 & 0.0175 & 11.263 & 0.0189 & 0 & 0.033 & {--}  & 0.032 & {--}   \\
202 & 242.88756	& -27.44722 & 160.14 & 1.76 & 74.60  & 137.84 & 11.697 & 0.0178 & 11.482 & 0.0196 & 1 & 0.033 & {--}  & {--}  & {--}   \\
203 & 242.90406	& -27.38113 & 131.84 & 3.07 & 36.80  & 127.11 & 12.418 & 0.0183 & 12.145 & 0.0207 & 0 & 0.069 & {--}  & 0.061 & {--}   \\
204 & 243.04386	& -27.35514 & 108.18 & 0.61 & 104.85 & 150.16 & 11.529 & 0.0175 & 11.333 & 0.0190 & 0 & 0.036 & {--}  & 0.027 & {--}   \\
209 & 242.46282	& -27.37831 & 147.53 & 2.85 & 48.96  & 123.48 & 12.173 & 0.0180 & 11.905 & 0.0201 & 0 & 0.068 & {--}  & 0.061 & {--}   \\
210 & 242.85333	& -26.92947 & 138.00 & 5.12 & 21.56  & 46.31  & 13.655 & 0.0207 & 13.413 & 0.0271 & 0 & 0.007 & 0.025 & 0.002 & 0.009  \\
\multicolumn{16}{c}{\dots} \\
\bottomrule
\end{tabular}
\tablefoot{The entire table is available in electronic form at the CDS.
\textbf{Columns:}
(4) Distances from Gaia DR3.
(5--6) Mass estimates assuming 5 and 10\:Myr from \cite{MiretRoig2022a}.
(7--10) {\it unWISE} W1--W2 magnitudes and associated uncertainties.
(11) Flag indicating possible contamination in {\it unWISE} images discussed in Sect.~\ref{sect:photometry} (1 = flagged, and 0 = clean).
(12--13) Excess probability inferred from analysis assuming 5 and 10\:Myr, respectively, keeping {\it unWISE}-flagged sources.
(14--15) Same as (10--11), but computed excluding {\it unWISE}-flagged sources.
}
\end{table}
\vspace*{\fill}
\end{landscape}

\onecolumn

\renewcommand{\arraystretch}{1.4}
\begin{table*}[!t]
\caption{Disk fractions estimated per mass bin assuming ages of 5 and 10\:Myr.}
\label{tab:fractions}
\centering
\begin{tabular}{
l
r@{\,--\,}l
l
c
l
c
}
\toprule
&
\multicolumn{2}{c}{{Flagged sources :}} &
\multicolumn{2}{c}{{\textbf{Excluded}}} &
\multicolumn{2}{c}{{\textbf{Included}}} \\
\cmidrule{2-7}
{Age} &
\multicolumn{2}{c}{{Mass Range}} &
{Fraction} &
{$N_\mathrm{obj}$} &
{Fraction} &
{$N_\mathrm{obj}$} \\
{(Myr)} &
\multicolumn{2}{c}{{(\Mj)}} &
{(\%)} &
&
{(\%)} &
\\ 
\midrule
\multirow{10}{*}{5 Myr}
 & 6     & 11.85   & $37.5\pm6.2$         & 32 & $38.0\pm6.0$          & 50 \\
 & 11.85 & 16.85   & $37.5_{-6.2}^{+3.1}$ & 32 & $46.0\pm4.0$          & 50 \\
 & 16.85 & 25      & $34.2\pm0.0$         & 38 & $36.7_{-0.0}^{+4.1}$  & 49 \\
 & 25    & 32.3    & $35.7_{-4.8}^{+2.4}$ & 42 & $37.0_{-2.2}^{+4.3}$  & 46 \\
 & 32.3  & 36      & $33.3_{-5.1}^{+2.6}$ & 39 & $30.4_{-2.2}^{+4.3}$  & 46 \\
 & 36    & 41.05   & $27.3_{-4.5}^{+2.3}$ & 44 & $28.3_{-2.2}^{+4.3}$  & 46 \\
 & 41.05 & 55      & $21.1\pm2.6$         & 38 & $26.1_{-2.2}^{+4.3}$  & 46 \\
 & 55    & 69.5    & $26.8_{-4.9}^{+2.4}$ & 41 & $26.5_{-2.0}^{+4.1}$  & 49 \\
 & 69.5  & 90      & $17.8_{-2.2}^{+4.4}$ & 41 & $18.0\pm2.0$          & 46 \\
 & 90    & 105     & $28.3_{-4.3}^{+2.2}$ & 45 & $28.8\pm3.4$          & 58 \\
\midrule
\multirow{8}{*}{10 Myr}
 & 8     & 13.7     & $38.5_{-7.7}^{+7.7}$ & 24 & $41.5_{-7.3}^{+7.3}$ & 39 \\
 & 13.7  & 15.55    & $25.9_{-0.0}^{+7.4}$ & 27 & $38.1_{-4.8}^{+7.1}$  & 42 \\
 & 15.55 & 25       & $32.1\pm3.6$         & 28 & $41.5\pm4.9$          & 41 \\
 & 25    & 48       & $32.4_{-0.0}^{+2.9}$ & 34 & $37.5_{-2.5}^{+5.0}$  & 40 \\
 & 48    & 63.1     & $30.6_{-0.0}^{+2.8}$ & 36 & $32.5\pm2.5$          & 40 \\
 & 63.1  & 77       & $27.0_{-0.0}^{+2.7}$ & 37 & $30.0_{-2.5}^{+5.0}$  & 40 \\
 & 77    & 94.3     & $16.7_{-0.0}^{+2.8}$ & 36 & $22.5\pm2.5$          & 40 \\
 & 94.3  & 105      & $15.2_{-2.2}^{+4.3}$ & 46 & $22.6\pm5.7$          & 53 \\
\bottomrule
\end{tabular}
\tablefoot{Results are shown both excluding and including sources flagged as potentially contaminated in {\it unWISE} images. This table is available in electronic form at the CDS.}
\end{table*}

\FloatBarrier
\section{Uncertainties}
\label{app:appendix}

\begin{figure*}[h!]
\begin{minipage}{0.5\textwidth}
    \centering
    \includegraphics[width=0.97\linewidth]{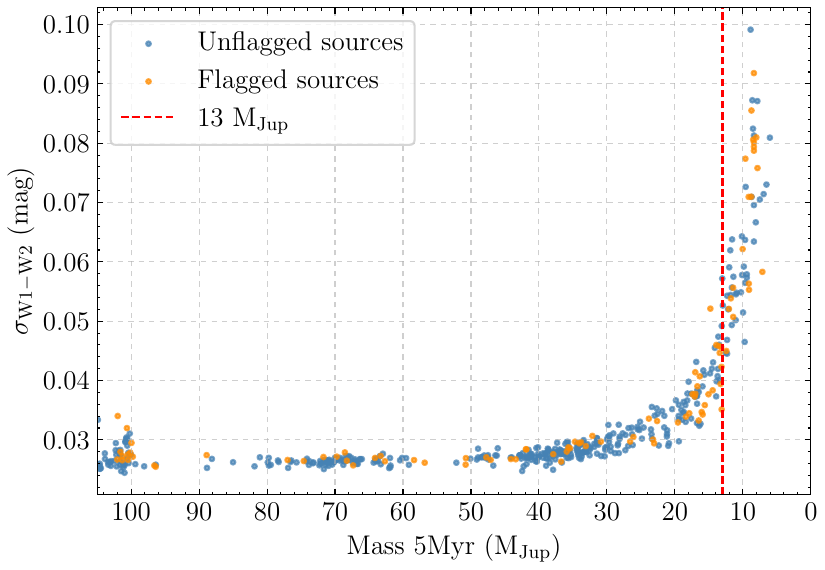}
\end{minipage}
\hfill
\begin{minipage}{0.5\textwidth}
    \centering
    \includegraphics[width=0.97\linewidth]{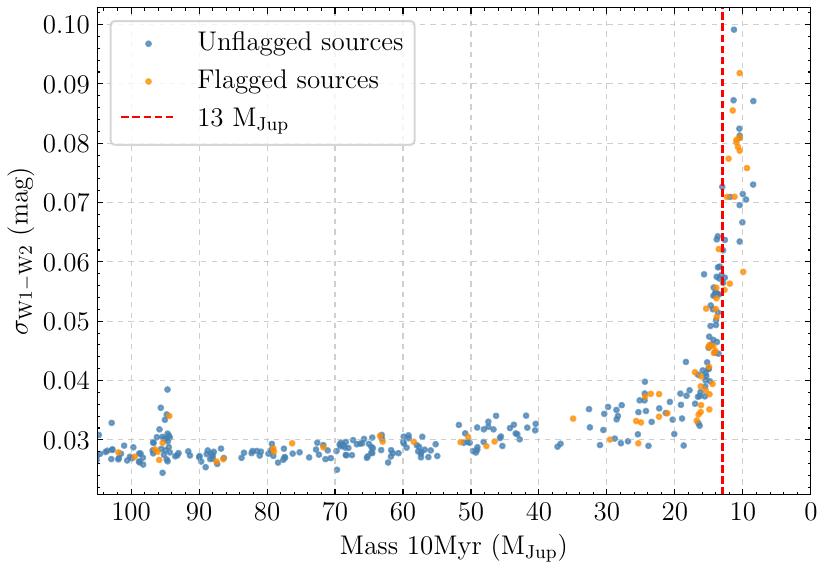}
\end{minipage}
    \caption{Photometric uncertainty on the (W1--W2) color as a function of mass at 5 ({\it Left}) and 10\:Myr ({\it Right}), for our sample described in Sect.~\ref{sect:photometry}. The uncertainties are computed as the quadratic sum of the W1 and W2 magnitude errors, i.e., $\sigma_{\mathrm{W1-W2}} = \sqrt{\sigma_{\mathrm{W1}}^2 + \sigma_{\mathrm{W2}}^2}$. Sources flagged as potentially blended in {\it WISE} images (see Sect.~\ref{sect:photometry}) are shown in orange.}
    \label{fig:errors}
\end{figure*}

\twocolumn

\end{appendix}

\end{document}